\documentclass[journal,a4paper]{IEEEtran}

\usepackage{latexsym}
\usepackage{graphicx}

\usepackage[mathscr]{eucal}

\usepackage{hyperref}

\usepackage{cite}
\usepackage{subcaption}
\usepackage{comment}
\usepackage{amsthm}
\usepackage{overpic}
\usepackage{steinmetz}
\usepackage{array}
\usepackage{url}
\usepackage{xcolor}
\IEEEoverridecommandlockouts

\theoremstyle{plain}

\usepackage{latexsym}
\usepackage{graphicx}
\usepackage{amsfonts,amssymb,amsmath}
\usepackage{varwidth}
\usepackage{hyperref}

\usepackage[T1]{fontenc}
\usepackage{cite}
\usepackage{subcaption}
\usepackage{comment}
\usepackage{overpic}
\usepackage{steinmetz}
\usepackage{array}
\usepackage{url}
\usepackage{xcolor}
\usepackage{algorithm,algorithmic}

\IEEEoverridecommandlockouts

\begin{document}

\title{{Beam-Coherence-Aware Two-Stage Digital Combining for mmWave MU-MIMO Systems}
}

\author{\IEEEauthorblockN{\normalsize
    Yasaman Khorsandmanesh, \textit{Student Member, IEEE},  Emil Björnson, \textit{Fellow, IEEE}, \\ Joakim Jaldén, \textit{Senior Member, IEEE}, and  Bengt Lindoff, \textit{Senior Member, IEEE}}
    \thanks{
Y.\ Khorsandmanesh, E.\ Björnson, and J.\ Jaldén are with the School of Electrical Engineering and Computer Science, KTH Royal Institute of Technology, Stockholm, Sweden (E-mails: \{yasamank, emilbjo, jalden\}@kth.se). B.\ Lindoff is with BeammWave AB (E-mail: bengt@beammwave.com).}
\thanks{{This work was supported by the Smarter Electronics System program by Vinnova and Vinnova through the SweWIN center (2023-00572).}}
\thanks{This work was presented in part at the IEEE 36th International Symposium on Personal, Indoor and Mobile Radio Communications (PIMRC 2025), 2025, Istanbul, Türkiye, which appears in this manuscript as reference \cite{khorsandmanesh2025channel}.}
}

\maketitle
{
\begin{abstract}
This paper considers a wideband millimeter-wave MIMO system with fully digital transceivers at both the base station and the user equipment (UE), focusing on mobile scenarios. To reduce the baseband processing burden at the UE, we propose a two-stage digital combining architecture, where the received signals are compressed from $K$ antennas to dimension $N_{\mathrm c}$ before baseband processing. The first-stage combining matrix exploits channel geometry and is updated on the beam-coherence timescale, which is longer than the channel coherence time, while the second stage is updated per channel coherence time. We develop a pilot-based channel estimation framework tailored to the proposed two-stage digital combining architecture, leveraging maximum likelihood estimation. Furthermore, we propose a time-domain method that exploits the finite delay spread to reconstruct the full channel from a reduced number of pilot subcarriers. Precoding and combining schemes are designed accordingly, and spectral efficiency expressions with imperfect channel state information are derived. Numerical results show that the proposed time-domain approach outperforms hybrid beamforming while reducing pilot overhead. We further demonstrate that the framework extends to multi-user MIMO and retains its performance advantages. These results highlight the potential of two-stage fully digital transceivers for future wideband systems.
\end{abstract}
}

{

\section{Introduction}

Millimeter-wave (mmWave) technology holds great potential for wireless networks due to its significantly larger bandwidth compared to sub-6 GHz systems \cite{rangan2014millimeter}. However, mmWave systems have faced deployment challenges in 5G. Key limitations include poor penetration through obstacles like walls, which limits the coverage in urban environments, and environmental factors like rain and fog that heavily attenuate mmWave signals, complicating service reliability \cite{uwaechia2020comprehensive}. Furthermore, effective beamforming for mmWave currently requires complex, power-intensive hardware, leading to rapid battery drain at the user equipment (UE) \cite{dutta2019case}. These limitations suggest that the widespread mmWave deployment will wait until significant advancements in hardware design, power management, and signal processing have occurred.

To overcome the severe path loss at these frequencies, systems employ large antenna arrays and multiple-input multiple-output (MIMO) techniques to achieve high beamforming gains \cite{bogale2014beamforming}. While hybrid beamforming (HBF) has been widely studied as a practical solution to reduce hardware complexity, fully digital beamforming (DBF) is increasingly attractive due to its flexibility and ability to fully exploit the spatial degrees of freedom offered by MIMO technology \cite{yang2018digital}. The main challenge with DBF is the high hardware and processing complexity. In particular, a UE with many antennas must process high-dimensional signals across many subcarriers, leading to substantial power consumption and data movement, especially in wideband systems \cite{bjornson2017massive, alkhateeb2014channel}. 

Channel estimation is another key challenge in mmWave systems \cite{hassan2020channel}. Conventional frequency-domain approaches estimate each subcarrier independently, leading to pilot overhead and computational complexity proportional to the number of subcarriers. However, OFDM channels exhibit inherent structure and can often be represented by a limited number of significant time-domain taps \cite{li2002simplified}. By exploiting this property, more efficient estimation can be developed that jointly processes information across subcarriers, thereby reducing both pilot overhead and computational complexity while improving estimation accuracy. 

In this paper, we propose a fully digital wideband mmWave point-to-point MIMO architecture that addresses these challenges. The UE employs a two-stage digital combining structure, where the downlink received signal is first compressed from $K$ UE receive antennas to $N_{\mathrm c}$ dimensions, with $N_{\mathrm s} \leq N_{\mathrm c} \leq K$, where $N_{\mathrm s}$ denotes the number of data streams. The first-stage combining is updated on the beam-coherence timescale, while the second stage adapts to small-scale fading per coherence block. The beam coherence time is defined as the duration over which beams remain aligned \cite{khorsandmanesh2024beam}. We also propose a time-domain channel estimation method that reduces pilot overhead by estimating the channel taps directly and reconstructing the full frequency-domain channel via discrete Fourier transforms.

While the framework is developed for single-user point-to-point MIMO (SU-MIMO) to clearly expose the design, it naturally extends to multi-user MIMO (MU-MIMO). In the multi-user case, inter-user interference is handled via linear precoding, and we adopt minimum mean-squared error (MMSE)-based designs while retaining the proposed two-stage combining structure at the UE.

\subsection{Related Works}

HBF has been extensively studied to reduce RF-chain requirements in mmWave systems \cite{heath2016overview}, but its analog constraints limit performance in wideband and dynamic scenarios. Fully digital architectures enable per-subcarrier processing and fine-grained spatial control, but introduce significant computational and hardware challenges \cite{lindoff2021ultimate}. Dimension-reduction and subspace-based methods have been proposed to mitigate these issues \cite{brady2013beamspace}, although they typically assume frequent updates or full-dimensional channel state information (CSI).

Fully digital architectures have recently gained attention due to advances in hardware design \cite{lindoff2021ultimate}. These architectures enable per-subcarrier beamforming and fine-grained spatial processing, which are particularly beneficial in wideband systems. However, the associated computational burden and data movement remain major challenges, especially at the UE side. DBF and HBF require a similar number of RF components \cite[Ch.~7]{bjornson2024introduction}, thus, the basic premise for the HBF is that phase shifters have lower power and cost than ADCs \cite{BeammWavewhite}.
However, if DBF is implemented with a lower ADC resolution, it can provide better energy efficiency \cite{roth2018comparison}. Several works have proposed dimension-reduction techniques or subspace-based processing to alleviate this issue, but often assume frequent updates of the combining matrices or full-dimensional channel state information \cite{alkhateeb2014channel,brady2013beamspace}.

Channel estimation in wideband MIMO systems is commonly performed in the frequency domain \cite{bjornson2024introduction}, which leads to high pilot overhead. Time-domain and structured estimation approaches exploit the finite delay spread to reduce the number of unknowns, but their integration with practical UE architectures is less explored. In \cite{venugopal2017time},  a sparse time-domain channel estimation method for hybrid mmWave systems is proposed that reduces training overhead by exploiting channel sparsity, whereas our approach leverages the finite delay spread to reconstruct the full channel with a simple DFT-based method and reduced pilot overhead without requiring sparsity assumptions. In \cite{kim2019two}, a low-complexity two-step time-domain channel estimation is proposed for hybrid mmWave systems that exploits angular and delay sparsity.

In MU-MIMO, linear precoding schemes such as minimum mean-square error (MMSE) precoding are commonly used to manage inter-user interference \cite{bjornson2017massive}. These methods require accurate channel state information and are typically studied under fully digital architectures without considering UE-side dimensionality reduction.

\subsection{Contributions}

This paper proposes a beam-coherence-aware two-stage digital combining architecture for wideband mmWave MIMO systems. The key idea is to reduce the received signal dimension before baseband processing while maintaining the flexibility of fully digital beamforming. The main contributions are:
\begin{itemize} \item We propose a beam-coherence-aware two-stage digital combining architecture for mmWave MIMO UEs, where the first-stage combining matrix is updated once per beam coherence time $\mathsf{T}{\mathrm B}$, while the second-stage combiner is updated once per channel coherence time $\mathsf{T}{\mathrm C}$.  \item We develop a pilot-based channel estimation framework for this architecture in both uplink and downlink. In addition to a conventional frequency-domain maximum likelihood (ML) estimator, we propose a time-domain-based estimation method that exploits the finite delay spread of the OFDM channel. This reduces the required pilot overhead from $S$ pilot subcarriers to $L$ pilot subcarriers, where $L\ll S$. \item We propose precoding and combining schemes tailored to the considered architecture. In the single-user case, the design is based on the singular value decomposition of the estimated channels, while in the multi-user extension, we employ MMSE precoding based on the estimated effective channels. \item We derive achievable spectral-efficiency (SE) expressions under imperfect CSI and quantify the performance of the proposed architecture. This includes both the single-user case and a multi-user extension based on a UatF bound. 
\end{itemize}

\subsection{Notation}
The sets of integer, real, and complex numbers are denoted by $\mathbb{Z}$, $\mathbb{R}$, and $\mathbb{C}$, respectively. Matrices and vectors are denoted by bold uppercase and lowercase letters, respectively, such as $\mathbf{X}$ and $\mathbf{x}$. The $(m,k)$th element of $\mathbf{X}$ is denoted by $[\mathbf{X}]_{m,k}$, while $[\mathbf{x}]_m$ denotes the $m$th element of $\mathbf{x}$. The identity matrix of size $M\times M$ is denoted by $\mathbf{I}_M$, and $\mathbf{0}_{M\times K}$ denotes the all-zero matrix of size $M\times K$. $\mathbf{X}_{(:,1:d)}$ denotes the submatrix formed by the first $d$ columns of $\mathbf{X}$, i.e., all rows and columns $1$ to $d$. The transpose, conjugate, conjugate transpose, pseudo-inverse, trace, and vectorization operators are denoted by $(\cdot)^{\mathrm T}$, $(\cdot)^*$, $(\cdot)^{\mathrm H}$, $(\cdot)^\dagger$, $\mathrm{tr}(\cdot)$, and $\mathrm{vec}(\cdot)$, respectively. The Frobenius norm and Euclidean norm are denoted by $\|\cdot\|_{\mathrm F}$ and $\|\cdot\|_2$, respectively.  Moreover, $\mathbb{E}\{\cdot\}$ denotes expectation and $\mathcal{CN}(0,\sigma^2)$ denotes a circularly symmetric complex Gaussian random variable with variance $\sigma^2$.

\subsection{Paper Outline}
The remainder of this paper is organized as follows. Section~\ref{sec:sys} presents the system model and the considered two-stage digital combining architecture. Section~\ref{sec:est} introduces the pilot-based channel estimation procedure, while Section~\ref{sec:TD-estimation} presents the proposed time-domain-based channel estimation method. Section~\ref{secCapacity} evaluates the achievable downlink spectral efficiency and describes the corresponding precoding and combining design. Section~\ref{sec:MUMIMO} extends the proposed framework to the MU-MIMO case. Section~\ref{sec:complexity_overhead} compares the pilot overhead and computational complexity of the proposed time-domain estimator with conventional frequency-domain estimation. Finally, Section~\ref{sec:numerical} provides numerical results and Section~\ref{sec6} concludes the paper.

}

\begin{figure}[!t]
    \centering
\includegraphics[scale = 0.35]{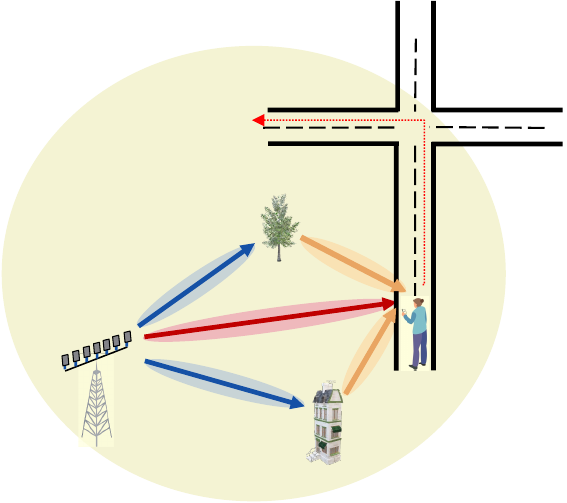}
    \caption{A mmWave MIMO mobile system.}
\label{fig:systemmodel}
\end{figure}

\section{System Model} \label{sec:sys}
\begin{figure}[t!]
  \centering
   \begin{overpic}[width=0.8\columnwidth]{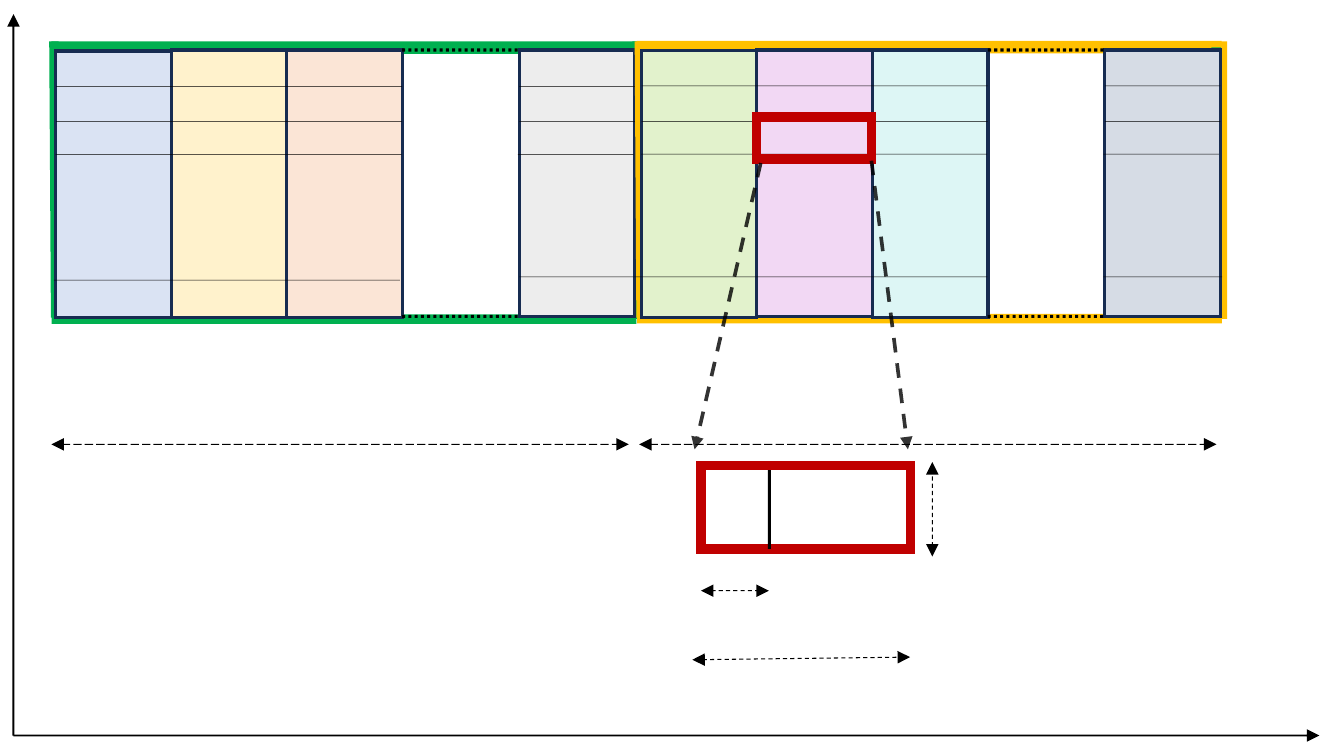}
   \put(4,54){\tiny $\tau =1$}%
   \put(13,54){\tiny $\tau =2$}%
   \put(22,54){\tiny $\tau =3$}%
   \put(32,54){\tiny $\cdots$}%
   \put(39,54){\tiny $\tau =\mathfrak{t}$}%
  \put(58,3){\scriptsize $\mathsf{T}_\mathrm{C}$}%
  \put(8,38){\scriptsize $\vdots$}%
  \put(17,38){\scriptsize $\vdots$}%
  \put(26,38){\scriptsize $\vdots$}%
  \put(32,39){\scriptsize $\cdots$}%
  \put(43,38){\scriptsize $\vdots$}%
  \put(22,19){\scriptsize $\mathsf{T}_\mathrm{B}$}%
  \put(71,17){\scriptsize $\nu$-th subcarrier}%
   \put(99,1){\footnotesize Time}
  \put(0,58){\footnotesize Frequency}%
   \put(54,9){\scriptsize $\mathsf{T}_{\rm p}$}%
   \end{overpic}
   \vspace{-1mm}
\caption{The channel is approximately time-invariant in each block $\tau$, comprising all subcarriers $S$ and channel coherence time $\mathsf{T}_\mathrm{C}$. The compressed digital combining matrix must be updated at the larger time intervals called the beam coherence time $\mathsf{T}_\mathrm{B}$, which is $\mathfrak{t}$ times larger than  $\mathsf{T}_\mathrm{C}$. $\mathsf{T}_{\mathrm{p}}$ denotes the pilot time, corresponding to the duration used for pilot transmission within a coherence interval. }
\label{fig:timeblock}
\vspace{-5mm}
\end{figure}

We first consider a wideband mmWave SU-MIMO system with $M$ antennas at the base station (BS) and $K$ antennas at the UE. { In Section~\ref{sec:MUMIMO}, we extend the results for MU-MIMO setups.} As depicted in Fig.~\ref{fig:systemmodel}, we focus on a mobile system in which the scheduled UE moves along a trajectory, while multi-user scenarios will be considered in future work. The propagation environment is modeled geometrically using  $N_\text{cl}$ scattering clusters \cite{rappaport2015millimeter}. We assume an OFDM signal with $S$ subcarriers and $L$ time-domain taps. The channel matrix undergoes continuous variations over time, which we model piecewise constantly. As depicted in Fig.~\ref{fig:timeblock}, we consider time-domain blocks numbered by $\tau$ that contain all $S$ subcarriers and have a time duration that matches the coherence time $\mathsf{T}_\mathrm{C}$ of the channel \cite{bjornson2017massive}. The frequency domain channel matrix for the $\nu$-th  subcarrier in the $\tau$-th block is denoted as \cite[Ch.~7]{bjornson2024introduction}
\begin{equation}
 \mathbf{H}[\tau,\nu] = \sum_{i=0}^{N_{\text{cl}}} \Big( \underbrace{\sum_{\ell=0}^{L-1} {\alpha_i}[\tau,\ell] e^{-j2\pi \ell v/S}}_{\Bar{\alpha}_i[\tau,\nu]} \Big) {\mathbf{a}_{\mathrm{r}}} ({\phi_{i}^{\mathrm{r}}}[\tau]) \mathbf{a}_{\mathrm{t}}^{\mathrm{T}} ({\phi_{i}^{\mathrm{t}}}[\tau]),   \label{eq:OFDM_ChannelModel}
\end{equation}
for  $\nu = 0,\ldots, S-1$. Here, ${\alpha_i}[\tau,\ell]\thicksim \mathcal{CN}(0, {\beta_{i}}[\tau,\ell])$ is the small-scale fading coefficient of the $\ell$-th time-domain tap for $\ell = 1,\ldots, L-1$, and $\beta_{i}[\tau,\ell] \triangleq \mathbb{E}\{ |\alpha_i[\tau,\ell]|^2 \}$ denotes the average power from the $i$-th cluster in the $\ell$-th tap.  $\beta_{i}[\tau,\ell]$ will vary gradually from block $\tau$ to other blocks. The line-of-sight path is denoted by $i = 0$ in \eqref{eq:OFDM_ChannelModel}, where ${\alpha}_0[\tau,0] = \sqrt{\beta_{0}}$, ${\alpha}_0[\tau,\ell] = 0$ for $\ell = 1,\ldots, L-1$, and $\beta_{0}$ describes the large-scale fading.
The vectors $\mathbf{a}_{\mathrm{r}}(\phi_{i}^{\mathrm{r}}[\tau])$ and $\mathbf{a}_{\mathrm{t}} (\phi_{i}^{\mathrm{t}}[\tau])$ are the array response vectors at the UE and BS, respectively. Both the UE and the BS employ horizontal uniform linear arrays (ULA) configuration with antenna spacing $\delta$ so that\footnote{This assumption is made to make the notation tractable, but can be easily extended to uniform planar arrays or even non-uniform array geometries.}  \cite{bjornson2024introduction}
\begin{align}
\mathbf{a}_{\mathrm{r}}(\phi_{i}^{\mathrm{r}}[\tau]) & = [1, e^{j 2\pi \delta \mathrm{sin}(\phi_{i}^{\mathrm{r}}[\tau]) / \lambda_\mathrm{c}},\ldots, e^{j 2\pi \delta(K-1) \mathrm{sin}(\phi_{i}^{\mathrm{r}}[\tau]) / \lambda_\mathrm{c}}]^{\mathrm{T}}, \label{eq:ar}  \\ 
\mathbf{a}_{\mathrm{t}} (\phi_{i}^{\mathrm{t}}[\tau]) & 
=[1, e^{j 2\pi \delta \mathrm{sin}(\phi_{i}^{\mathrm{t}}[\tau])/\lambda_\mathrm{c}},\ldots, e^{j 2 \pi  \delta (M-1) \mathrm{sin}(\phi_{i}^{\mathrm{t}}[\tau])/\lambda_\mathrm{c}}]^{\mathrm{T}} ,  \label{eq:at} 
\end{align} 
where $\phi_{i}^{\mathrm{r}}[\tau]$ and $\phi_{i}^{\mathrm{t}}[\tau]$ denotes the azimuth angle of arrival (AoA) and the azimuth angle of departure (AoD) measured from the broadside direction of the respective arrays in the block $\tau$, and $\lambda_\mathrm{c}$ is the wavelength at the carrier frequency $f_\mathrm{c}$.

The channel matrix in \eqref{eq:OFDM_ChannelModel} changes continuously over time. The parameters $\Bar{\alpha}_i[\tau,\nu]$ undergoes rapid fluctuations, while the AoA $\phi_{i}^{\mathrm{r}}[\tau]$ and the AoD $\phi_{i}^{\mathrm{t}}[\tau]$, and also $\beta_{i}[\tau,\ell]$, evolve slowly as they are determined by the large-scale geometry. The beam coherence time $\mathsf{T}_\mathrm{B}$ was defined in \cite{khorsandmanesh2024beam} as the duration over which the angular directions remain approximately fixed from a beamforming perspective, so one can keep the beamforming vectors constant. As indicated in Fig.~\ref{fig:timeblock}, $\mathsf{T}_\mathrm{B}$, is much larger than $\mathsf{T}_\mathrm{C}$ ($\mathfrak{t}=\mathsf{T}_\mathrm{B}/\mathsf{T}_\mathrm{C}$ times larger, shown by the green box). The beam coherence time is typically at least one order of magnitude larger than the channel coherence time, although both scale with factors such as UE mobility \cite{khorsandmanesh2024beam}.

    \begin{figure*}
        \centering
        \subfloat[Proposed setup]{
   \begin{overpic}[scale = 0.42]{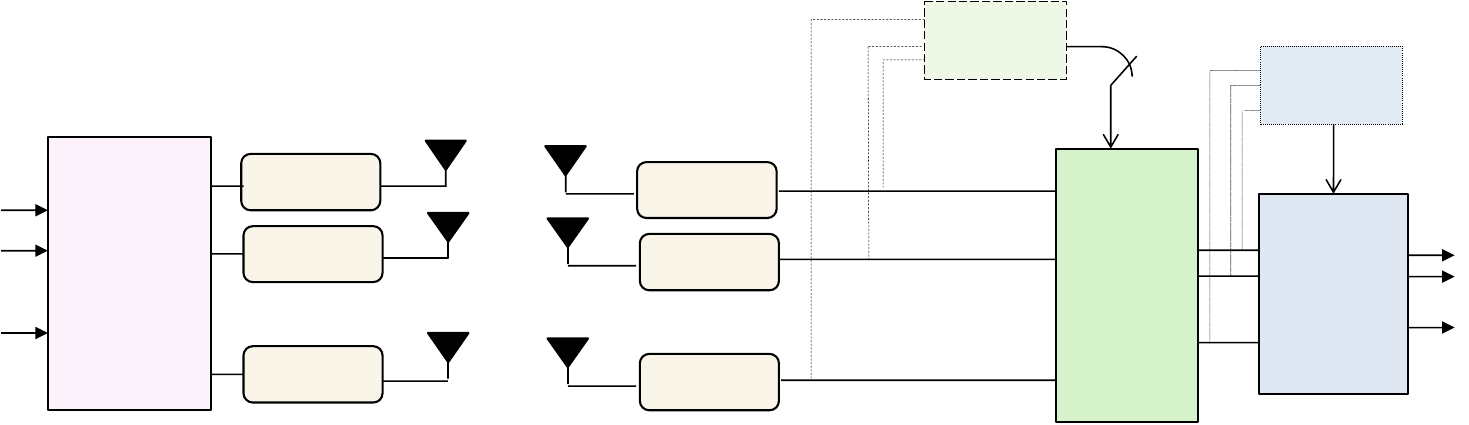}
\put(-1,14.5){\tiny $1$}
\put(-1,12){\tiny $2$}
\put(0.5,7.5){\tiny $\vdots$}
\put(-2.5,6.5){\tiny $N_\mathrm{s}$}
\put(5,15){\tiny  Digital}
\put(5,12.5){\tiny  Precoding}
\put(5,10){\tiny  $\mathbf{F}[\tau,\nu]$}
\put(32,19.5){\tiny $1$}
\put(32,14.5){\tiny $2$}
\put(30,7.5){\tiny $\vdots$}
\put(31,6.5){\tiny $M$}
\put(32,11){\tiny mmWave}
\put(32,9){\tiny Channel}
\put(36.5,19.5){\tiny $1$}
\put(36.5,14.5){\tiny $2$}
\put(39,7.5){\tiny $\vdots$}
\put(36.5,6.5){\tiny $K$}
\put(18,16){\tiny  RF Chain}
\put(18,11){\tiny  RF Chain}
\put(18,3){\tiny  RF Chain}
\put(45,15.5){\tiny  RF Chain}
\put(45,10.5){\tiny  RF Chain}
\put(45,2.5){\tiny  RF Chain}
\put(84.5,12.5){\tiny $1$}
\put(84,10.5){\tiny $2$}
\put(84.5,7){\tiny $\vdots$}
\put(82.5,6.5){\tiny $N_\mathrm{c}$}
\put(65.5,27){\tiny  CSI of }
\put(64,25){\tiny $K \times N_\mathrm{c}$}
\put(76,28){\tiny Occasionally}
\put(77,26){\tiny updated}
\put(73,16){\tiny  First-Stage}
\put(73,14){\tiny Digital}
\put(73,12){\tiny Combining}
\put(73,10){\tiny (Dimension}
\put(73,8){\tiny Reduction)}
\put(73,6){\tiny $\mathbf{Q}[\nu]$}
\put(88.5,23.5){\tiny  CSI of }
\put(86.5,21.5){\tiny  $N_\mathrm{c} \times N_\mathrm{s}$}
\put(87,12){\tiny Baseband}
\put(87,10){\tiny Digital}
\put(87,8){\tiny Combining}
\put(87,6){\tiny $\mathbf{W}[\tau,\nu]$}
\put(99.5,12){\tiny $1$}
\put(99.5,10){\tiny $2$}
\put(97,7){\tiny $\vdots$}
\put(99,6.5){\tiny $N_\mathrm{s}$}
\end{overpic}
            \label{subfig:A}
        }\quad
        \subfloat[Implementation of BeammWave's setup]{
            \includegraphics[width=.3\linewidth]{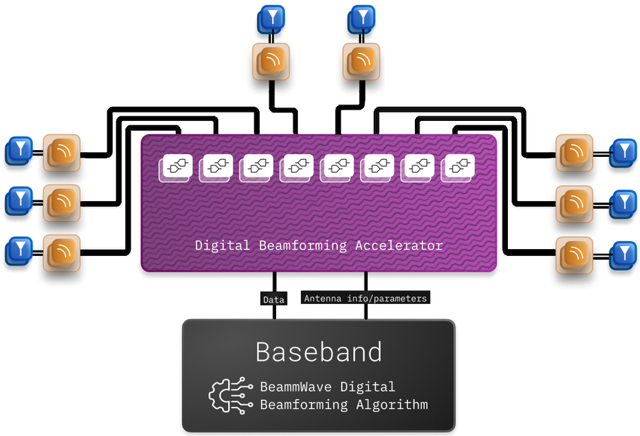}
            \label{subfig:B}
        }
        \caption{Block diagram of a mmWave SU-MIMO system employing fully digital precoding and a two-stage digital combining architecture. a) In the proposed setup, the first-stage combining matrix $\mathbf{Q}[\nu]$ reduces the signal dimension and occasionally accesses CSI across all $K$ antennas. The second-stage combining  $\mathbf{W}[\tau,\nu]$ is updated frequently but has a reduced dimension $N_\mathrm{c} < K$. b) The BeammWave company utilizes the same structure for implementing its UE receiver chips.}
        \label{fig:hardware}
    \end{figure*}
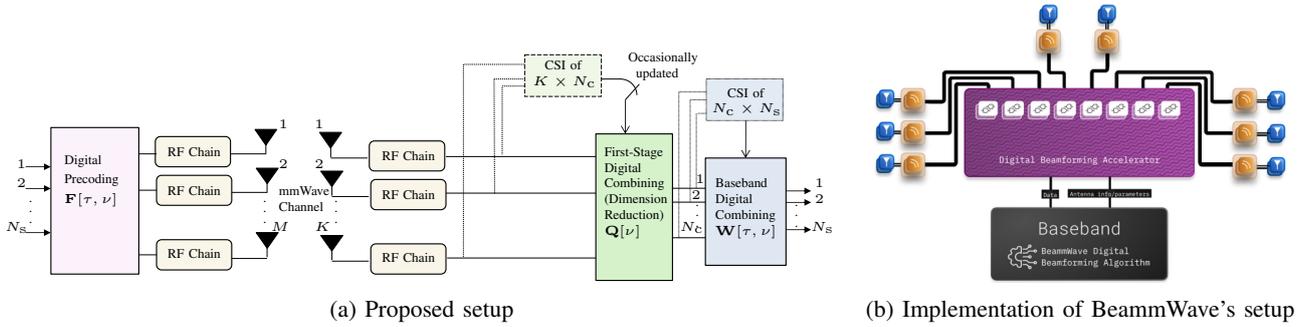

\subsection{Hardware Architecture Properties} \label{sec:hardware}

Even if there has been much prior work on HBF, the long-term goal is to implement DBF in mmWave systems, as its digital processing over each antenna element enables fast and flexible adaptation to channel variations \cite{lindoff2021ultimate}. This is possible using state-of-the-art transceiver technology, but the bottleneck is the vast amount of baseband samples that must be processed. A potential solution is to use the UE architecture shown in Fig.~\ref{subfig:A}, which contains $K$ RF chains and a first-stage digital combining that reduces its dimension to $N_\mathrm{c}$, where $N_\mathrm{s} \le N_\mathrm{c}\le K$, so the rest of the baseband processing (i.e., channel estimation, second-stage combining) can be done with a similar dimensionality as in current HBF methods. Here, $N_\mathrm{s}$ is the number of data streams.
The key to success is that the first-stage combining is implemented efficiently and updated infrequently, so full-dimensional CSI is only required occasionally. Fig.~\ref{subfig:B} shows BeammWave company's DBF platform on which the proposed algorithm could be implemented \cite{beammwave2023platform}.

Based on the beam coherence time concept, it is logical that the first-stage digital combining is updated once per $\mathsf{T}_\mathrm{B}$, while it is fixed within one green block in Fig.~\ref{fig:timeblock}. 
Since the small-scale fading varies more rapidly (i.e., once per coherence time), both the digital precoding $\mathbf{F}[\tau,\nu]$ and second-stage baseband combining $\mathbf{W}[\tau,\nu]$ are updated at this interval for each $\tau$ and $\nu$-th subcarrier.  In contrast, the first-stage digital combining $\mathbf{Q}[\nu]$ remains constant over $\mathfrak{t}$ coherence blocks.

A key difference between the proposed two-stage digital architecture and conventional HBF is that the first-stage digital combining can vary arbitrarily over the subcarriers. By contrast, the analog combining in a hybrid architecture also reduces the dimension but is the same on all subcarriers, since it is implemented using phase shifters.
In the remainder of the paper, we will explore how to operate the considered two-stage digital architecture when it comes to uplink and downlink channel estimation and what rates are achieved. Note that we assume that the BS uses DBF without hardware limitations to focus on the design of low-complexity UEs.

\vspace{3mm}
\section{Frequency-Domain Channel Estimation}
\label{sec:est}

In this section, we describe how to acquire CSI in the system shown in Fig.~\ref{fig:hardware}. To set the digital precoding matrix $\mathbf{F}[\tau,\nu]$ and the two digital combining matrices $\mathbf{Q}[\nu]$ and $\mathbf{W}[\tau,\nu]$, both the BS and UE need to estimate the channel $\mathbf{H}[\tau,\nu]$. We consider time-division duplexing (TDD) operation, where the channel is first estimated in the uplink and used in the downlink by leveraging channel reciprocity. The first block ($\tau=1$) in the green box in Fig.~\ref{fig:timeblock} is handled differently from the others, as the first-stage combining $\mathbf{Q}[\nu]$ is selected and then remains fixed until the end of the green box. We then estimate the reduced-dimension effective channel $\mathbf{Q}^\mathrm{H}[\nu]\mathbf{H}[\tau,\nu]$ for $\tau >1$. Thus, we consider  $\tau=1$ and $\tau >1$ separately in the remainder of this section. We need uplink and downlink pilots as we have multiple antennas at both the BS and UE sides.

\subsection{Channel Estimation in the First Block: $\tau=1$} \label{sec:channelestimate}

In the first block of the $\mathfrak{t}$ that fits in a beam coherence time, the UE transmits pilot sequences on each subcarrier, which the BS uses to estimate the complete channel. Each antenna at the UE transmits an orthonormal pilot sequence of length $\mathsf{t}_{\mathrm{p}} \geq K$. The pilot overhead can be expressed as the ratio $\mathsf{t}_{\mathrm{p}}/\mathsf{t}_\mathrm{c}$, where $\mathsf{t}_\mathrm{c}$ indicates coherence block length in symbols. 

The pilot matrix used by the UE is denoted as $\boldsymbol{\Phi}^{\text{U}}[1,\nu] \in \mathbb{C}^{  K \times  \mathsf{t}_{\mathrm{p}}}$ and has orthonormal rows. The UE transmits $\sqrt{\mathsf{t}_{\mathrm{p}}} \boldsymbol{\Phi}^{\text{U}}[1,\nu]$ to ensure that the total pilot energy is proportional to the pilot length. The received signal at the BS is 
\begin{equation}
\boldsymbol{\mathbf{Y}}^{\text{Pilot,U}}[1,\nu]= \sqrt{P_\mathrm{r}  \mathsf{t}_{\mathrm{p}}} \mathbf{H}^\mathrm{T}[1,\nu] \boldsymbol{\boldsymbol{\Phi}}^{\text{U}}[1,\nu] + \boldsymbol{\mathbf{N}}^{\text{Pilot,U}}[1,\nu], \label{eq:pilotup1} 
\end{equation}
where $P_\mathrm{r} $ is the transmit power used by the UE, normalized by the noise power, and $\boldsymbol{\mathbf{N}}^{\text{Pilot,U}}[1,\nu] \in \mathbb{C}^{M \times \mathsf{t}_{\mathrm{p}}}$ is the noise matrix at the BS with i.i.d. $\mathcal{CN}(0,1)$-entries.

Many channel estimators can be developed based on the received pilot signal in \eqref{eq:pilotup1}. In this paper, we adopt the classical ML estimation approach since we consider a mobile UE where prior statistical channel knowledge cannot be obtained. The ML estimate of $\mathbf{H}^\mathrm{T}[1,\nu]$ based on $\boldsymbol{\mathbf{Y}}^{\text{Pilot,U}}[1,\nu]$ is \cite{Kay1993a} 
\begin{equation}  \label{eq:channelMLuplink} \hat{\mathbf{H}}^\mathrm{T}[1,\nu] = \boldsymbol{\mathbf{Y}}^{\text{Pilot,U}}[1,\nu]\boldsymbol{\Phi}^{\text{U}^\dagger}[1,\nu]/ \sqrt{P_\mathrm{r}\mathsf{t}_\mathrm{p}},
\end{equation}
where $(\cdot)^\dagger$   denotes the pseudo-inverse of a matrix. We have ${\boldsymbol{\boldsymbol{\Phi}}}^{\text{U}^\dagger}[1,\nu]= {\boldsymbol{\boldsymbol{\Phi}}}^{\text{U}^\mathrm{H}}[1,\nu] ({\boldsymbol{\boldsymbol{\Phi}}}^{\text{U}}[1,\nu]{\boldsymbol{\boldsymbol{\Phi}}}^{\text{U}^\mathrm{H}}[1,\nu])^{-1} $ in this case.

Now, we shift focus to the downlink data transmission. The received signal on the $\nu$-th subcarrier through $\mathbf{H}[1,\nu]$ for the first block is given as 
\begin{align}
\mathbf{Y}^{\text{D}}[1,\nu] &= \underbrace{{\mathbf{H}}[1,\nu]  \mathbf{F}[1,\nu]}_{\mathbf{B}[1,\nu]}  \mathbf{S}[1,\nu] +  \mathbf{N}^{\text{D}}[1,\nu],  \label{eq:recievedsignal}
\end{align}
where $\mathbf{S}[1,\nu] \in \mathbb{C} ^{N_\mathrm{s} \times N_\mathrm{d}}$ represents the symbol matrix with independent entries that have unit norms, $N_\mathrm{s} \le \text{rank}(\mathbf{H}[1,\nu])$ is the number of spatially multiplexed data streams transmitted on each subcarrier, and $N_\mathrm{d}$ denotes the number of symbol vectors $\mathbf{s}[1,\nu]\in \mathbb{C} ^{N_\mathrm{s} \times 1}$ (one column of the symbol matrix) that are sent in the downlink. Each entry of $\mathbf{N}^{\text{D}}[1,\nu] \in \mathbb{C}^{ K \times N_\mathrm{d}}$ is i.i.d. $ \mathcal{CN}(0,1)$. The digital precoding $\mathbf{F}[1,\nu] \in \mathbb{C}^{ M \times N_\mathrm{s}}$ is set by the BS utilizing the channel estimates obtained from the uplink in \eqref{eq:channelMLuplink}. Moreover,
$ \Vert \mathbf{F}[1,\nu] \Vert ^2_\mathrm{F}= P_\mathrm{t}$ represents the per-subcarrier transmit
power normalized by
the noise power, where $\Vert \cdot  \Vert_{\mathrm{F}}$ denotes the Frobenius norm. Section~\ref{se:SVD} provides details on how to select the precoding matrix $\mathbf{F}[1,\nu]$.

The notation ${\mathbf{B}}[1,\nu] = {\mathbf{H}}[1,\nu]  \mathbf{F}[1,\nu]$ will be used for the precoded channel. A viable method to provide the UE with CSI is to transmit pilots using the precoding matrix, so the UE can estimate ${\mathbf{B}}[1,\nu]$. The downlink pilot matrix is $\boldsymbol{{\Phi}}^{\text{D}}[1,\nu] \in \mathbb{C}^{N_{\mathrm{s}} \times N_{\mathrm{s}}}$. By transmitting these pilots over the channel in \eqref{eq:recievedsignal}, the received signal at the UE becomes
\begin{equation}
{\mathbf{Y}}^{\text{Pilot,D}}[1,\nu]= \sqrt{ N_{\mathrm{s}}} {\mathbf{B}[1,\nu]}{\boldsymbol{\boldsymbol{\Phi}}}^{\text{D}}[1,\nu] + {\mathbf{N}}^\text{Pilot,D}[1,\nu],
\end{equation} 
where $\boldsymbol{\mathbf{N}}^{\text{Pilot,D}}[1,\nu] \in \mathbb{C}^{K \times {N}_{\mathrm{s}}}$ is the noise matrix  with i.i.d. $\mathcal{CN}(0,1)$-entries. 
By following the same steps as in the uplink, the estimated precoded channel at $\tau=1$ becomes
\begin{equation}
\hat{\mathbf{B}}[1,\nu] = {\mathbf{Y}}^{\text{Pilot,D}}[1,\nu] {\boldsymbol{\boldsymbol{\Phi}}}^{{\text{D}}^\dagger}[1,\nu] /\sqrt{N_{\mathrm{s}}}.
    \label{eq:estimateB1}
\end{equation}
Now, the UE can select the first-stage digital combining matrix ${\mathbf{Q}}[\nu] \in \mathbb{C}^{K \times N_\mathrm{c}}$ and lower-dimensional second-stage digital combining matrix ${\mathbf{W}}[1,\nu]\in \mathbb{C}^{N_\mathrm{c} \times K}$ based on  $\hat{\mathbf{B}}[1,\nu] $ and ${\mathbf{Q}}^\mathrm{H}[\nu]\hat{\mathbf{B}}[1,\nu]$, respectively. In Section~\ref{se:SVD}, we explain how to select them and why we need the second combining matrix.

\subsection{Channel Estimation for Blocks  $\tau>1$}

For $ 1<\tau \le \mathfrak{t}$, we need to repeat the steps of uplink channel estimation and estimation of the precoded downlink channel, but now the first-stage combining matrix $\mathbf{Q}[\nu]$ is fixed. We define the effective channel ${\mathbf{G}} [\tau,\nu] = \mathbf{Q} ^\mathrm{H}[\nu]\mathbf{H}[\tau,\nu] \in \mathbb{C}^{ N_\mathrm{c} \times M}$. The UE sends the orthonormal pilot matrix $\grave{\boldsymbol{{\Phi}}}^{\text{U}}[\tau,\nu] \in \mathbb{C}^{{N}_{\mathrm{c}} \times \mathsf{t}_{\mathrm{p}}}$ to estimate ${\mathbf{G}} [\tau,\nu]$. The received signal at the BS is  
\begin{equation}\grave{\boldsymbol{\mathbf{Y}}}^{\text{Pilot,U}}[\tau,\nu]= \sqrt{P_\mathrm{r}  \mathsf{t}_{\mathrm{p}}} \mathbf{G}^\mathrm{T}[\tau,\nu] \grave{\boldsymbol{\Phi}}^{\text{U}}[\tau,\nu] + \grave{\boldsymbol{\mathbf{N}}}^{\text{Pilot,U}}[\tau,\nu], \label{eq:pilotup} 
\end{equation}
which is similar to \eqref{eq:pilotup1} but contains the lower-dimensional effective channel. Here, $\grave{\boldsymbol{\mathbf{N}}}^{\text{Pilot,U}}[\tau,\nu] \in \mathbb{C}^{M \times \mathsf{t}_{\mathrm{p}}}$ is the noise matrix with i.i.d. $\mathcal{CN}(0,1)$-entries. The ML estimate of $\mathbf{G}[\tau,\nu]$ is
\begin{equation}
  \label{eq:estimategt}  \hat{\mathbf{G}}^\mathrm{T}[\tau,\nu] = \grave{\mathbf{Y}}^{\text{Pilot,U}}[\tau, \nu] \grave{\boldsymbol{\Phi}}^{\text{U}^\dagger}[\tau,\nu] /\sqrt{ P_\mathrm{r}\mathsf{t}_{\mathrm{p}}}.
\end{equation}

The BS selects the precoding $\mathbf{F}[\tau,\nu]$ based on $\hat{\mathbf{G}}[\tau,\nu]$ (see Section~\ref{se:SVD} for details). At the UE, the received signal after the first-stage digital combining matrix is
\begin{align}
\grave{\mathbf{Y}}^{\text{D}}[\tau,\nu]  & = \underbrace{{\mathbf{G}}[\tau,\nu]\mathbf{F}[\tau,\nu]}_{{\mathbf{D}} [\tau,\nu]}\mathbf{S}[\tau,\nu] + \underbrace{\mathbf{Q} ^\mathrm{H}[\nu]{\mathbf{N}}^{\text{D}}[\tau,\nu]}_{\grave{\mathbf{N}}^{\text{D}}[\tau,\nu]},  \label{eq:secendeffective}
\end{align}
where ${\mathbf{D}} [\tau,\nu] = {\mathbf{G}}[\tau,\nu]\mathbf{F}[\tau,\nu]$ is the precoded effective channel. Since the first-stage combining matrix $\mathbf{Q}[\nu]$ only reduces the channel dimension based on the channel geometry but is independent of the current small-scale fading realizations, we also need a second-stage combining matrix ${\mathbf{W}}[\tau,\nu]$ to mitigate interference between the $N_\mathrm{s}$ streams. We need to estimate ${\mathbf{D}} [\tau,\nu]$ to implement that.

Now, by transmitting the downlink pilot matrix $\grave{\boldsymbol{\Phi}}^{\text{D}}[\tau,\nu] \in \mathbb{C}^{N_{\mathrm{s}} \times N_{\mathrm{s}}  }$, the received signal after first-stage  combining matrix is
\begin{equation} \label{eq:pilot-correlated}
\grave{\mathbf{Y}}^{\text{Pilot,D}}[\tau, \nu]= \sqrt{  N_{\mathrm{s}}} {\mathbf{D}}[\tau,\nu]\grave{\boldsymbol{\Phi}}^{\text{D}}[\tau,\nu] + \grave{\mathbf{N}}^\mathrm{Pilot,D}[\tau,\nu],
\end{equation}
where $\grave{\mathbf{N}}^\mathrm{Pilot,D}[\tau,\nu] \in \mathbb{C}^{N_{\mathrm{c}} \times N_{\mathrm{s}} }$ is the colored processed noise with independent columns and the covariance matrix $\mathbf{C}_{\grave{\mathbf{N}}^\mathrm{Pilot,D}} = \mathbb{E} \left[ \mathbf{Q}^\mathrm{H}[\nu]\mathbf{Q}[\nu] \right]$.  
Since the pilot matrix is multiplied from the right in \eqref{eq:pilot-correlated} and the noise correlation appears from the left, the correlation has no impact on the estimator. 

The ML estimate of effective downlink channel $\mathbf{D}[\tau,\nu]$ is 
\begin{equation}
\hat{\mathbf{D}}[\tau,\nu] = \frac{ \grave{\mathbf{Y}}^{\text{Pilot,D}}[\tau, \nu] \grave{\boldsymbol{\Phi}}^{\text{D}^{\dagger}}[\tau,\nu]}{\sqrt{ N_{\mathrm{s}}}}. \label{eq:estimateD}  
\end{equation}
All the proposed steps are presented in Algorithm \ref{Alg:ML_Estimation}.

 \begin{algorithm}
 [t!] 
 \small \caption{Proposed Channel Estimation Approach} \label{Alg:ML_Estimation} \begin{algorithmic}[1] \STATE{\textbf{Input}: Pilot matrices $\boldsymbol{\Phi}^{\text{U}}[1,\nu]$, $\boldsymbol{{\Phi}}^{\text{D}}[1,\nu]$, $\grave{\boldsymbol{\Phi}}^{\text{U}}[\tau,\nu]$,  $\grave{\boldsymbol{{\Phi}}}^{\text{D}}[\tau,\nu]$}   \IF{$\tau = 1$}     \STATE{Estimate uplink channel $    {\mathbf{H}}^\mathrm{T}[1,\nu]$ by utilizing \eqref{eq:channelMLuplink}}    
 \STATE{Set precoding $\mathbf{F}[1,\nu]$ as in \eqref{eq:firstF}}
 \STATE{Estimate downlink channel after precoding $    {\mathbf{B}}[1,\nu]$ as in \eqref{eq:estimateB1}} 
  \STATE{Set compressed first-stage combining matrix  $\mathbf{Q}[\nu]$ as in \eqref{eq:firstQ}}
    \STATE{Set second-stage digital combining matrix $\mathbf{W}[1,\nu]$ based on lower-dimensional channel $\mathbf{Q}[\nu]\hat{\mathbf{B}}[1,\nu]$ as in \eqref{eq:firstW}}
        
 \ELSIF{$1<\tau \le \mathfrak{t}$}
 \STATE{Define effective channel  ${\mathbf{G}}[\tau,\nu] = {\mathbf{Q}}^\mathrm{H}[\nu]{\mathbf{H}}[\tau,\nu]$} 
blue
\STATE {Estimate uplink effective channel ${\mathbf{G}}^\mathrm{T}[\tau,\nu]$ as in \eqref{eq:estimategt}} \STATE{Update precoding $\mathbf{F}[\tau,\nu]$  as in \eqref{eq:F}}
\STATE {Estimate second effective channel ${\mathbf{D}}[\tau,\nu] $ as in \eqref{eq:estimateD}}
 \STATE{Update second-stage digital combining matrix  $\mathbf{W}[\tau,\nu]$ as in \eqref{eq:w}}
 \ENDIF 
\STATE \textbf{Output}: $\mathbf{F}[\tau,\nu]$, $\mathbf{Q}[\nu]$ and $\mathbf{W}[\tau,\nu]$     
\end{algorithmic}
\end{algorithm}

{
\section{Time-Domain-Based Channel Estimation}
\label{sec:TD-estimation}

In Section~\ref{sec:est}, we described a frequency-domain channel estimation approach for TDD systems, where the channel is estimated independently on each subcarrier, as summarized in Algorithm~\ref{Alg:ML_Estimation}. While straightforward, this approach has several limitations. First, it requires pilots on all subcarriers, which leads to a large pilot overhead that reduces the achievable SE. Second, the computational complexity scales linearly with the number of subcarriers, leading to increased latency and power consumption. Third, frequency-domain estimation is sensitive to
noise and interference, often yielding noisy channel estimates unless additional smoothing or interpolation is applied. Most importantly, this approach fails to exploit the inherent structure of OFDM-based channels, as it treats the channel responses on different subcarriers as independent unknowns, and thereby ignoring the strong correlation across frequency induced by the finite delay spread of the channel.

These limitations motivate a time-domain-based estimation approach. The key observation is that the OFDM channel is fully characterized by a finite number of time-domain taps $L$. Instead of estimating the channel on each subcarrier, we estimate these taps directly. Since the number of taps is typically much smaller than the number of subcarriers ($L \ll S$), the number of unknown parameters is significantly reduced. For example, in our simulations in Section \ref{sec:numerical} with $S=512$ and $L=6$, we have $S/L \approx 86$, so the cyclic prefix is about 1.17\%. These limitations motivate a time-domain-based estimation approach. The key observation is that the OFDM channel is fully characterized by a finite number of time-domain taps $L$. Instead of estimating the channel on each subcarrier $\nu$, we estimate these taps directly. Since the number of taps is typically much smaller than the number of subcarriers ($L \ll S$), the number of unknown parameters is significantly reduced. For example, in our simulations in Section \ref{sec:numerical} with $S=512$ and $L=6$, we have $S/L \approx 86$, which corresponds to a cyclic prefix overhead of only  1.17\% of the symbol duration. This enables joint processing across subcarriers and reduces the number of unknown parameters by exploiting the finite delay spread. Consequently, the pilot overhead can be significantly reduced by estimating a smaller set of channel parameters while still enabling reconstruction of the channel over the entire bandwidth.

\subsection{Channel Representation}
We focus on the uplink channel estimation problem. Let's start with first block $\tau=1$ and the uplink channel matrix $\mathbf{H}^{\mathrm T}[1,\nu] = [\mathbf{h}_1[1,\nu], \ldots, \mathbf{h}_{K}[1,\nu]] \in \mathbb{C}^{M \times K}$, where $\mathbf{h}_k[1,\nu] \in \mathbb{C}^{M}$ denotes the channel vector associated with the $k$-th UE transmit antenna where $k=1,\ldots,K$ during the first transmission block $\tau=1$. The received uplink signal corresponding to the $k$-th UE on subcarrier $\nu$ is expressed as
\begin{equation}
\mathbf{y}_k^{\rm U}[1,\nu]
= \sqrt{P_\mathrm{r}} 
\mathbf{h}_k[1,\nu] x[1,\nu]
+
\mathbf{n}_k^{\rm U}[1,\nu],
\quad \nu=0,\ldots,S-1 ,
\end{equation}
where $x[1,\nu]$ is a pilot symbol and $\mathbf{n}_k^{\rm U}[1,\nu]\in\mathbb{C}^{M}$ is additive noise with i.i.d.\ $\mathcal{CN}(0,1)$ entries at the BS. For notational simplicity, we assume that the pilot sequence length is $\mathsf{t}_{\mathrm{p}} = 1$ per UE $k$.

The frequency-domain channel vector $\mathbf{h}_k[1,\nu]\in\mathbb{C}^{M}$ can be expressed as
\begin{equation}
\mathbf{h}_k[1,\nu]
=
\sum_{\ell=0}^{L-1}
\bar{\mathbf{h}}_k^{\mathrm T}[1,\ell]
e^{-j2\pi \ell \nu / S},
\label{eq:td_fd_relation}
\end{equation}
where $\bar{\mathbf{h}}_k[1,\ell]$ denotes the $\ell$-th time-domain channel tap and $L$ is the effective channel length. Equation~\eqref{eq:td_fd_relation} shows that the frequency-domain channel is the $S$-point discrete Fourier transform (DFT) of the time-domain channel taps.

Collecting the frequency-domain channel vectors across all subcarriers yields
\begin{equation}
\boldsymbol{\mathcal{H}}_k
\triangleq
\big[
\mathbf{h}_k[1,0], \ldots, \mathbf{h}_k[1,S-1] 
\big] \in \mathbb{C}^{M \times S},
\end{equation}
and we define $\boldsymbol{\mathcal{H}}=[\boldsymbol{\mathcal{H}}_1,\ldots,\boldsymbol{\mathcal{H}}_{K}] \in \mathbb{C}^{M \times SK}$.

Assuming unit-modulus pilot symbols, i.e., $x_k[1,\nu]=1$ for all $\nu$ and $P_\mathrm{r} $ the transmit power used by the UE, the received signals across subcarriers can be stacked as in \eqref{eq:time-freq}  on the top of the next page,
\begin{figure*}[t!]
{
 \begin{align}
\underbrace{\Big[{\mathbf{y}}^{\rm U}_k[1,0], \dots, {\mathbf{y}}_k^{\rm U}[1,S-1] \Big]}_{\overset{\Delta}{=} \boldsymbol{\mathcal{Y}}_k} = \underbrace {\boldsymbol{\mathcal{H}}_k }_{[\bar{\boldsymbol{\mathcal{H}}}_k, \mathbf{0}_{(L+1)\times(S-L-1)}]\sqrt{P_\mathrm{r}S}\textbf{DFT}_S .1 }  +   \underbrace{[{\mathbf{n}}^{\rm U}_k[1,0], \dots, {\mathbf{n}}^{\rm U}_k[1,S-1]]}_{\overset{\Delta}{=} \boldsymbol{\mathcal{N}}_k} \label{eq:time-freq}
\end{align} }
\end{figure*}
\begin{figure*}
{
 \begin{equation}
[{\mathbf{h}}_k[1,\nu_0], \dots, {\mathbf{h}}_k[1,\nu_{L-1}]]
= [\bar{\mathbf{h}}_k[1,0], \dots, \bar{\mathbf{h}}_k[1,L-1]]
\underbrace{\begin{bmatrix}
1 & e^{-j2\pi\frac{\nu_0}{S}} & \dots & e^{-j2\pi\frac{\nu_{L-1} }{S}} \\
\vdots & \vdots & \ddots & \vdots \\
1 & e^{-j2\pi\frac{\nu_0(L-1)}{S}} & \dots & e^{-j2\pi\frac{\nu_{L-1}(L-1)}{S}}
\end{bmatrix}}_{\textbf{DFT}_L}  \label{eq:frequency-time-channels} 
\end{equation}  }
\end{figure*}
where $\textbf{DFT}_S$ denotes the $S\times S$ DFT matrix. Applying the inverse DFT yields
\begin{equation}
\frac{1}{\sqrt{P_\mathrm{r}S}}\boldsymbol{\mathcal{Y}}_k\cdot \textbf{DFT}_S^{-1} = [\bar{\boldsymbol{\mathcal{H}}}_k,\mathbf{0}] + \frac{1}{\sqrt{P_\mathrm{r}S}}\boldsymbol{\mathcal{N}}_k\cdot \mathbf{DFT}_S^{-1},
\end{equation}
where the last $S-L$ columns correspond to zero padding.

\subsection{Time-Frequency Conversion}

Instead of transmitting pilots on all $S$ subcarriers, the $L$ time-domain channel taps can be uniquely recovered from only $L$ pilot subcarriers $\{\nu_0,\ldots,\nu_{L-1}\}$. This allows us to express the channel responses jointly across these subcarriers, leading to \eqref{eq:frequency-time-channels}, which establishes a linear transformation between the $L$ frequency-domain observations and the $L$ time-domain channel taps.

A convenient and well-conditioned choice is to select equally spaced pilot subcarriers\footnote{This formulation applies when $\frac{S}{L}$ is an integer, which enables uniform spacing of the selected pilot subcarriers. If this condition is not satisfied, one can approximate the spacing by rounding to the nearest integer, selecting the corresponding subcarriers $S$, and discarding any excess subcarriers that are not needed ($\left\lceil \frac{S}{L} \right\rceil\times L-S$). This results in $ \frac{S}{L} $ distinct pilot patterns, each enabling full channel estimation with identical overhead. The pilot subcarriers can also be selected with an offset $\delta \in \left\{0,\ldots,\left\lceil \tfrac{S}{L} \right\rceil - 1\right\}$, given by $\nu_\ell
=
\delta + S\frac{\ell}{L}$ for $
\ell=0,\ldots,L-1$.}
\begin{equation}
\nu_\ell
=
S\frac{\ell}{L},
\qquad
\ell=0,\ldots,L-1,
\label{eq:uniform_pilots}
\end{equation}

We first select $L$ pilot subcarriers according to~\eqref{eq:uniform_pilots} and transmit pilots $x[1,\nu_\ell]=P_\mathrm{r} /K$ on each of them.  The ML estimate of $\mathbf{h}_k[1,\nu_\ell]$ is given by
\begin{equation}
\hat{\mathbf{h}}_k[1,\nu_\ell]
=
\mathbf{y}_k^{\rm U}[1,\nu_\ell],
\quad \ell=0,\ldots,L-1,\; k=1,\ldots,K.
\end{equation}

Using the relation in~\eqref{eq:frequency-time-channels}, these estimated frequency-domain responses are converted to the time domain by applying the inverse $L$-point DFT, which yields the estimates of the $L$ time-domain channel taps $\hat{\bar{\mathbf{h}}}_k[1,\ell]$.

\subsection{Frequency-Domain Channel Reconstruction}

Once the $L$ time-domain channel taps have been estimated, the channel across all subcarriers can be reconstructed. Let
\begin{equation}
\hat{\bar{\boldsymbol{\mathcal{H}}}}_k
=
\big[
\hat{\bar{\mathbf{h}}}_k[1,0],\ldots,
\hat{\bar{\mathbf{h}}}_k[1,L-1]
\big]
\end{equation}
denote the estimated time-domain channel taps. To obtain the full frequency-domain channel, the vector is first zero-padded to length $S$ as
\begin{equation}
\hat{\boldsymbol{\mathcal{H}}}_{k,\mathrm{TD}}
=
\big[
\hat{\bar{\boldsymbol{\mathcal{H}}}}_k,
\mathbf{0}_{M\times(S-L)}
\big].
\end{equation}

The frequency-domain channels on all $S$ subcarriers are then obtained by applying the $S$-point DFT
\begin{equation}
\hat{\boldsymbol{\mathcal{H}}}_k
=
\frac{1}{\sqrt{S}}
\hat{\boldsymbol{\mathcal{H}}}_{k,\mathrm{TD}}
\mathbf{DFT}_S .
\end{equation}

This procedure reconstructs the full frequency-domain channel from only $L$ pilot subcarriers. Consequently, for $\mathsf{t}_p=1$, the pilot overhead is reduced from $S$ symbols in conventional frequency-domain estimation to only $L$ symbols.

Algorithm~\ref{Alg:TD_Estimation} summarizes the proposed time-domain-based channel estimation procedure. The same time-domain estimation principle can also be extended to the effective channels $\mathbf{B}[\tau,\nu]$, $\mathbf{G}[\tau,\nu]$ and $\mathbf{D}[\tau,\nu]$ introduced in Section~\ref{sec:est}. However, unlike $\mathbf{H}^{\mathrm T}[1,\nu]$, these effective channels may exhibit additional frequency selectivity due to subcarrier-dependent precoding and combining matrices inside them. To limit this effect, the $S$ subcarriers are partitioned into $N_{\mathrm{sub}}$ subbands, each containing
\begin{equation}
S_{\mathrm{sub}} = \frac{S}{N_{\mathrm{sub}}}
\end{equation}
subcarriers. Within each subband, the precoding and combining matrices are assumed to be approximately constant, that is,
\begin{equation}
\mathbf{F}_k[\tau,\nu] \approx \mathbf{F}_k[\tau,b],
\quad
\mathbf{W}_k[\tau,\nu] \approx \mathbf{W}_k[\tau,b],
\end{equation}
for all subcarriers $\nu$ belonging to subband $b=1,\ldots,N_{\mathrm{sub}}$. This approximation ensures that the effective channel within each subband remains governed by a limited delay spread $L_{\mathrm{eff}}$, which allows the time-domain estimation method to be applied using only $L_{\mathrm{eff}} \geq L$ pilot subcarriers per subband.  The time-domain estimation method is then applied using $L_{\mathrm{eff}}$ pilot subcarriers, followed by reconstruction across all subcarriers.

\begin{algorithm}[t]
\small

\caption{{Time-Domain-Based Channel Estimation}}
\label{Alg:TD_Estimation}
\begin{algorithmic}[1] {
\STATE \textbf{Input:} $S$, $L$, $\delta$, and $\{\mathbf{y}_k^{\rm U}[1,\nu_\ell]\}$
\STATE Select pilot subcarriers $\nu_\ell=\delta+\frac{S\ell}{L}$, $\ell=0,\ldots,L-1$
\STATE Transmit $x[1,\nu_\ell]=1$
\FOR{$k=1,\ldots,K$}
    \STATE Estimate pilot-subcarrier channels:
    $
    \hat{\mathbf{h}}_k[1,\nu_\ell]
    =
    \mathbf{y}_k^{\rm U}[1,\nu_\ell]
    $
    \STATE Form $\hat{\mathbf{H}}_{k,\mathrm{FD}}=[\hat{\mathbf{h}}_k[1,\nu_0],\ldots,\hat{\mathbf{h}}_k[1,\nu_{L-1}]]^{\mathrm T}$
    \STATE Compute taps:
    $
    \hat{\bar{\mathbf{H}}}_{k,\mathrm{TD}}=\mathbf{DFT}_L^{-1}\hat{\mathbf{H}}_{k,\mathrm{FD}}
    $
    \STATE Zero-pad to length $S$
    \STATE Reconstruct full channel:
    $
    \hat{\boldsymbol{\mathcal{H}}}_k=\frac{1}{\sqrt{S}}\mathbf{DFT}_S\hat{\boldsymbol{\mathcal{H}}}_{k,\mathrm{TD}}
    $
\ENDFOR
\STATE \textbf{Output:} $\{\hat{\boldsymbol{\mathcal{H}}}_k\}_{k=1}^{K}$}
\end{algorithmic}
\end{algorithm}
}

{\subsection{Complexity and Pilot Overhead Comparison}
\label{sec:complexity_overhead}

We next compare the proposed time-domain-based channel estimation approach with conventional frequency-domain estimation in terms of pilot overhead and computational complexity. Table~\ref{tab:complexity_overhead}  summarizes the key differences between the two approaches in terms of pilot overhead and computational complexity, which we will explain in the following. 

\subsubsection{Pilot Overhead}

In conventional frequency-domain channel estimation, the channel response is estimated independently on each of the $S$ subcarriers. As a result, $\mathsf{t}_{\mathrm p}\cdot S$ pilot symbols are required to estimate the channel over one transmission block, where $\mathsf{t}_{\mathrm p}$ denotes the number of pilot OFDM symbols. The total number of pilot symbols grows linearly with the system bandwidth, which increases the number of unknown channel coefficients and the estimation complexity. This can become prohibitive in wideband systems, even though the relative pilot overhead remains constant. The parameter $\mathsf{t}_{\mathrm p}$ must be selected such that the transmitted pilot sequences are orthogonal, which typically requires $\mathsf{t}_{\mathrm p} \geq K$ (or equivalently, the number of transmitted spatial streams). Hence, $\mathsf{t}_{\mathrm p}$ scales with the number of transmit antennas rather than the number of receive antennas. In contrast, the proposed time-domain-based method exploits the finite delay spread of the channel. Since the channel is fully characterized by $L$ time-domain taps with $L\ll S$, only $\mathsf{t}_{\mathrm p}\cdot L$ pilot symbols are sufficient to recover the full frequency-domain channel response. The resulting pilot overhead reduction factor is therefore
\begin{equation}
\frac{\text{Overhead}_{\text{TD}}}{\text{Overhead}_{\text{FD}}}
=
\frac{L}{S}.
\end{equation}
In practical OFDM systems, the ratio $\frac{L}{S}$ is kept small to limit the cyclic prefix overhead, meaning that the channel remains sparse relative to the total number of subcarriers.

\begin{table}[t]
\centering
\caption{ {Frequency- vs.\ Time-Domain Channel Estimation}}
\label{tab:complexity_overhead}
\begin{tabular}{p{2.6cm} p{2.3cm} p{2.3cm}}
\hline
\textbf{Metric} 
& \textbf{Freq.-Domain} 
& \textbf{Time-Domain} \\
\hline
Estimation domain 
& Per subcarrier 
& Channel taps \\
\hline
Unknown parameters 
& $KMS$ 
& $KML$ \\
\hline
Pilot symbols 
& $\mathsf{t}_{\mathrm p}S$ 
& $\mathsf{t}_{\mathrm p}L$ \\
\hline
Pilot scaling 
& $\mathcal{O}(S)$ 
& $\mathcal{O}(L)$ \\
\hline
Estimation complexity 
& $\mathcal{O}(KMS)$ 
& $\mathcal{O}(KML)$ \\
\hline
FFT operations 
& None 
& $L$-IFFT + $S$-FFT \\
\hline
FFT complexity 
& -- 
& $\mathcal{O}(L\log S)$ \\
\hline
Channel structure 
& Not exploited 
& Exploited \\
\hline
Noise averaging 
& No 
& Yes \\
\hline
Wideband suitability 
& Limited 
& High \\
\hline
\end{tabular}
\end{table}

\subsubsection{Computational Complexity}

Frequency-domain channel estimation requires processing each subcarrier independently. For each receive antenna, this involves $S$ complex-valued channel estimations, typically through least-squares or MMSE processing. The overall computational complexity scales as $\mathcal{O}(K M S)$, ignoring constant factors related to the estimator.

The proposed time-domain approach, on the other hand, performs channel estimation using only $L$ pilot subcarriers, followed by an inverse DFT of size $L$ and a DFT of size $S$ to reconstruct the frequency-domain channel. Since the estimated time-domain channel has only $L$ taps, it is zero-padded to length $S$ before applying the transform. Therefore, this corresponds to a partial DFT, in the sense that only $L$ non-zero coefficients contribute to the $S$ frequency-domain samples. The dominant operations per receive antenna are thus $\mathcal{O}(K M L) +\mathcal{O}(K L \log S)$, where the first term corresponds to estimating the $L$ time-domain taps and  the second term accounts for the FFT-based transformation to the frequency domain with only $L$ non-zero inputs. Since $L \ll S$ in typical wireless channels, the time-domain method substantially reduces the estimation complexity, particularly in the channel estimation stage, while the transformation to the frequency domain remains efficient due to the sparse (zero-padded) structure. The comparison reveals that the proposed time-domain-based channel estimation approach offers significant reductions in both pilot overhead and estimation complexity compared to conventional frequency-domain estimation. These gains become increasingly pronounced as the system bandwidth grows, making the proposed method particularly attractive for wideband and high-frequency systems where large numbers of subcarriers are employed.}
\section{Downlink Achievable SE}\label{secCapacity}
The proposed channel estimation procedure provides the BS and UE with CSI, which makes the selection of precoding and combining matrices feasible. The available CSI at the UE is crucial in determining the combining matrices and, ultimately, the achievable downlink SE. Therefore, in this section, we will explore the SE that can be achieved with two different levels of CSI availability at the UE: perfect and imperfect CSI.

The achievable SE is upper bounded by the channel capacity \cite{tse2005fundamentals}. In cases where the UE has imperfect CSI, the classical Shannon capacity formula cannot be evaluated directly. However, there are well-established lower bounds that can be utilized to characterize the SE under imperfect CSI. We will apply the \textit{use-and-then-forget} (UatF) technique \cite{bjornson2017massive}, where the UE uses the channel estimates to design the receive combining but disregards them in the signal detection process. In this section, we first consider the ideal scenario where the UE has perfect CSI and describe how to select the precoding and combining matrices in our setup. Then, 
we present an SE expression based on the UatF technique.

\subsection{Achievable SE with Perfect CSI at the UE}

We first consider the genie-aided case where the channel ${\mathbf{H}}[\tau,\nu]$ is assumed to be perfectly known to the UE. This is a benchmark for the practical estimation method proposed in this paper. With the transmission of Gaussian data symbols, the ergodic achievable
SE on the subcarrier $\nu$ is given by 
\begin{align}
     R[\nu] & =  \mathbb{E} \Bigg[ \log_2 \bigg(\det \Big( \mathbf{I}_{N_\mathrm{s}} +  \big({\mathbf{Q}}[\nu]{\mathbf{W}}[\tau,\nu] \big)^\dagger \mathbf{H}[\tau,\nu] \mathbf{F}[\tau,\nu] \nonumber \\ & \quad \times \mathbf{F}^\mathrm{H}[\tau,\nu]\mathbf{H}^\mathrm{H}[\tau,\nu] \big({\mathbf{Q}}[\nu]{\mathbf{W}}[\tau,\nu] \big) \Big) \bigg) \Bigg],
\end{align}
where the expectation is computed with respect to the fading process $\{ \mathbf{H}[\tau,\nu] \}$ \cite[Ch.~8]{tse2005fundamentals}. 
  The average SE over the subcarriers is
\begin{equation}
\text{SE}^{\text{full CSI}} =   \frac{1}{S} \sum_{\nu = 0}^{S-1}  \rho R[\nu],  \label{eq:capacityfullsytem}
\end{equation}
where $\rho = 1- \frac{\mathsf{t}_{\mathrm{p}} + N_\mathrm{s}}{\mathsf{t}_{\mathrm{c}}}$ compensates for the estimation overhead.

\subsection{Precoding and Combining Design} \label{se:SVD}

It remains to select the precoding and combining matrices to maximize the SE, for example, the expression in \eqref{eq:capacityfullsytem}. This is a classical problem with a well-established solution under perfect CSI \cite[Ch.~4]{bjornson2024introduction}. 
The optimal approach uses the singular value decomposition (SVD) of each subcarrier’s channel matrix to decouple the MIMO channels into many parallel channels. We take the same approach in our system, but consider the estimated channel matrix. For the $\nu$-th subcarrier and $\tau$-th block, the estimated channel matrix $\hat{\mathbf{H}}[\tau,\nu]$ can be decomposed via the SVD as
\begin{equation}
 \hat{\mathbf{H}}[\tau,\nu] = \mathbf{U}[\tau,\nu]\mathbf{\Lambda}[\tau,\nu]\mathbf{V}^\mathrm{H}[\tau,\nu],
\end{equation}
where ${\mathbf{U}}[\tau,\nu]  \in \mathbb{C}^{K \times K}$ is a unitary matrix containing the left singular vectors, ${\mathbf{\Lambda}}[\tau,\nu] \in \mathbb{C}^{K \times M}$ is a diagonal matrix containing the singular values in decreasing order on the diagonal, and ${\mathbf{V}}[\tau,\nu]  \in \mathbb{C}^{M \times M}$ is a unitary matrix containing the right singular vectors. The precoding matrix at the $\nu$-th subcarrier for $\tau=1$ is
\begin{equation}
    \mathbf{F}[1,\nu] =  \mathbf{V}_{(:,1:N_\mathrm{s})}[1,\nu]\operatorname{diag}(\sqrt{P_{1,\nu}}, \ldots,\sqrt{P_{N_\mathrm{s},\nu}}) , \label{eq:firstF} 
\end{equation}
where $\operatorname{diag}(\sqrt{P_{1,\nu}}, \ldots,\sqrt{P_{N_\mathrm{s},\nu}})$ is an $N_{\text{s}} \times N_{\text{s}}$ diagonal matrix with $P_{i,\nu}$ representing the $i$-th diagonal element. This matrix represents a power allocation over the $i$ streams and follows a water-filling strategy with the
total transmit power constraint $\Sigma^{N_\mathrm{s}}_{i=1} P_{i,\nu} = P_\mathrm{t} $ per subcarrier. Here, $\mathbf{V}_{(:,1:N_\mathrm{s})}[1,\nu] \in \mathbb{C}^{M \times N_\mathrm{s}}$ contains the first $N_\mathrm{s}$ columns of ${\mathbf{V}}[\tau,\nu]$, corresponding to the dominant $N_\mathrm{s}$ singular values.

To set the first-stage compressed combining matrix, the UE utilizes the SVD of the estimated precoded channel $\hat{\mathbf{B}}[1,\nu] = \mathbf{U}_{\mathrm{B}}[1,\nu]\mathbf{\Lambda}_{\mathrm{B}}[1,\nu]\mathbf{V}_{\mathrm{B}}^\mathrm{H}[1,\nu]$ and pick the first $N_\mathrm{c}$ columns of left singular matrix $\mathbf{U}_{\mathrm{B}}[1,\nu]$ as
\begin{equation}
    {\mathbf{Q}}[\nu] = \mathbf{U}_{\mathrm{B}(:,1:N_\mathrm{c})}[1,\nu]. \label{eq:firstQ}
\end{equation}
For block $\tau  =1$, we can set the second-stage combining as \begin{align}
  {\mathbf{W}}[1,\nu]  =
\begin{bmatrix}
\mathbf{I}_{N_{\mathrm s}} \\
\mathbf{0}_{(N_{\mathrm c}-N_{\mathrm s}) \times N_{\mathrm s}}
\end{bmatrix}, \label{eq:firstW}
\end{align}
which is non-square due to the SVD-based selection of ${\mathbf{Q}}[\nu]$. The most significant $N_\mathrm{s}$ signal components are already concentrated in the leading dimensions in \eqref{eq:firstQ}, making additional processing unnecessary at initialization. 

For $ 1<\tau \le \mathfrak{t}$, the first-stage combining matrix ${\mathbf{Q}}[\nu]$ remains fixed. The system uses the SVD of the estimated effective channel $\hat{\mathbf{G}}[\tau,\nu ] = \mathbf{U}_{\mathrm{G}}[\tau,\nu]\mathbf{\Lambda}_{\mathrm{G}}[\tau,\nu]\mathbf{V}_{\mathrm{G}}^\mathrm{H}[\tau,\nu]$ and selects the precoding as 
\begin{equation}
    \mathbf{F}[\tau,\nu] =  \mathbf{V}_{\mathrm{G}(:,1:N_\mathrm{s})}[\tau,\nu] \operatorname{diag}(\sqrt{P_{1,\nu}}, \ldots,\sqrt{P_{N_\mathrm{s},\nu}}),\label{eq:F} 
\end{equation}
using water-filling power allocation.
The second-stage combining is calculated based on estimated precoded effective channel $\hat{\mathbf{D}}[\tau,\nu] = \mathbf{U}_{\mathrm{D}}[\tau,\nu]\mathbf{\Lambda}_{\mathrm{D}}[\tau,\nu]\mathbf{V}_{\mathrm{D}}^\mathrm{H}[\tau,\nu]$ as
\begin{equation}
    {\mathbf{W}}[\tau,\nu] = \mathbf{U}_{\mathrm{D}(:,1:N_\mathrm{s})}[\tau,\nu]. \label{eq:w}
\end{equation}

\subsection{Achievable SE with Imperfect CSI at the UE}

We will now utilize the UatF approach to characterize the achievable ergodic with imperfect CSI. In this technique, the UE uses the channel estimate to compute the first- and second-stage combining matrices (as described in the last subsection), but 
treats the effective channel as a deterministic quantity when computing the SE. An arbitrary column of the received signal after the second-stage combining can be expressed as 
\begin{align}
& \breve{\mathbf{y}}[\tau,\nu]  = {\mathbf{W}^\mathrm{H}}[\tau,\nu]
{\mathbf{D}}[\tau,\nu]
\mathbf{s}[\tau,\nu] + {\mathbf{W}^\mathrm{H}}[\tau,\nu]   \grave{\mathbf{n}}[\tau,\nu] 
 \nonumber 
\\ &   = \bar{\mathbf{E}}[\nu] \mathbf{s}[\tau,\nu] + (\mathbf{E}[\tau,\nu] - \bar{\mathbf{E}}[\nu]) \mathbf{s}[\tau,\nu]   +  {\mathbf{W}^\mathrm{H}}[\tau,\nu]   \grave{\mathbf{n}}[\tau,\nu],
\label{eq:newnoise}
\end{align}
where $\mathbf{E}[\tau,\nu] = {\mathbf{W}}^\mathrm{H}[\tau,\nu]  {\mathbf{D}}[\tau,\nu] \in \mathbb{C}^{K \times K }$ represents the effective channel after applying the second receive combining matrix and 
$\bar{\mathbf{E}}[\nu]
$ denotes its mean with respect to the fading variations. 
This distinction between the effective channel and its mean is crucial for handling the statistical properties of the channel in subsequent analysis. The term $\Grave{\Grave{\mathbf{n}}}_d[\tau,\nu] = (\mathbf{E}[\tau,\nu] - \bar{\mathbf{E}}[\nu]) \mathbf{s}[\tau,\nu]       + {\mathbf{W}^\mathrm{H}}[\tau,\nu]   \grave{\mathbf{n}}[\tau,\nu]$ represents spatially colored noise term that 
is uncorrelated with the first term and has the covariance matrix
\begin{equation} \mathbf{C}_{\Grave{\Grave{\mathbf{n}}}_d} = \mathbb{E} \left[ \Grave{\Grave{\mathbf{n}}}_d[\tau,\nu]\Grave{\Grave{\mathbf{n}}}_d^\mathrm{H}[\tau,\nu] \right]. 
\end{equation}
By interpreting \eqref{eq:newnoise} as a  MIMO channel with the deterministic channel $\bar{\mathbf{E}}[\nu]$ and the colored noise $\Grave{\Grave{\mathbf{n}}}_d$, we can now write the average achievable SE of the system as 
\begin{equation}
\text{SE}^{\text{Imperfect CSI
 }} =   \frac{1}{S} \sum_{\nu = 0}^{S-1}  \rho R^{\text{Imperfect CSI}}[\nu] , \label{eq:imperfect}
\end{equation}
where $\rho = 1- \frac{\mathsf{t}_\mathrm{p} + N_\mathrm{s}}{\mathsf{t}_c}$ and the SE at the subcarrier $\nu$ is 
\begin{equation}
R^{\text{Imperfect CSI}}[\nu] =  \log_2 \left( \det \left( \mathbf{I}_{N_\mathrm{s}} + \bar{\mathbf{E}}^\mathrm{H}[\nu] \mathbf{C}_{\Grave{\Grave{\mathbf{n}}}_d}^{-1} \bar{\mathbf{E}}[\nu] \right) \right).
\end{equation}

\section{MU-MIMO Systems}
\label{sec:MUMIMO}

The previous sections considered a SU-MIMO system in order to present the proposed two-stage combining architecture and the associated channel estimation procedure with notational brevity. In this section, we briefly discuss how the framework extends to the multi-user case. To this end, consider a system where the BS serves $U$ UEs simultaneously. Each UE is equipped with $K$ antennas and employs the two-stage combining architecture described in Section~\ref{sec:hardware}. The BS is equipped with $M$ antennas and transmits $N_{\rm s}$ streams per UE, which is generally smaller than $K$.

The received signal at UE $u$ on subcarrier $\nu$ during block $\tau$ can be written as
\begin{equation}
\tilde{\mathbf y}_u[\tau,\nu]
=
\mathbf{ W}_u^{\mathrm H}\mathbf{Q}_u^{\mathrm H}[\nu]\mathbf{H}_u[\tau,\nu]\sum_{i=1}^{U}\mathbf{F}_i[\tau,\nu]\mathbf{x}_i[\tau,\nu]
+
\tilde{\mathbf n}_u[\tau,\nu],
\end{equation}
where $\tilde{\mathbf n}_u[\tau,\nu]= \mathbf{ W}_u^{\mathrm H}\mathbf Q_u^{\mathrm H}[\nu]\mathbf n_u[\tau,\nu] \in\mathbb{C}^{ K \times N_{\rm s}}$ is additive noise with i.i.d.\ $\mathcal{CN}(0,1)$ entries. $\mathbf{H}_u[\tau,\nu]\in\mathbb{C}^{K\times M}$ denote the downlink channel between the BS and UE $u$ on subcarrier $\nu$ during block $\tau$, $\mathbf{F}_i[\tau,\nu]\in\mathbb{C}^{M\times N_{\rm s}}$ is the precoding matrix for UE $i$, $\mathbf{x}_i[\tau,\nu]\in\mathbb{C}^{N_{\rm s}}$ is the arbitrary transmitted symbol vector, $\mathbf{Q}_u[\nu]\in\mathbb{C}^{K\times N_{\rm c}}$ is the first-stage combining matrix and $\mathbf{W}_u[\tau,\nu]\in\mathbb{C}^{N_{\rm c}\times K}$ is second-stage combining matrix.

\subsection{Channel Estimation}

The proposed time-domain channel estimation method in Section~\ref{sec:TD-estimation} can be applied independently for each UE. During uplink transmission, each UE $u$ transmits pilots on the selected $L$ pilot subcarriers. Note that to avoid pilot contamination, different UEs are assigned shifted versions of the pilot subcarriers (different $\delta$), which provides separation in the frequency domain. This avoids the need for orthogonal pilot sequences but reduces flexibility in pilot allocation.

\subsection{Precoding and Combining Design}
\label{sec:MU_precoding}

In MU-MIMO scenarios, the precoding must not only enhance the desired signal for each UE, but also suppress inter-user interference. For this reason, the SVD-based precoding used in SU-MIMO is no longer suitable, as it does not account for this. Instead, we follow a per-user linear MMSE precoding strategy based on the estimated effective channels. This choice follows the same design as the MMSE-based precoding matrix in~\cite{m2020mmse}.

For $\tau=1$, the MMSE precoding matrix is
\begin{equation}
\mathbf{F}_u[1,\nu]
=
\eta_u[1,\nu]
\left(
\hat{\mathbf{H}}_u^{\mathrm H}[1,\nu]\hat{\mathbf{H}}_u[1,\nu]
+
\mu \mathbf{I}_M
\right)^{-1}
\hat{\mathbf{H}}_u^{\mathrm H}[1,\nu],
\label{eq:MU_peruser_MMSE_precoder}
\end{equation}
where $\mu=\frac{U\sigma_n^2}{P_t}$ is a regularization parameter, $\sigma_n^2$ denotes the noise variance and $P_t$ is the total transmit power. The normalization factor $\eta_u[\tau,\nu]$ is selected such that the total power constraint for UE $u$  $\sum_{k=1}^{K} \sum_{i=1}^{N_{\rm s}}P_{k,i,\nu}=P_t$ per subcarrier satisfied. UE $u$ selects its first-stage combining matrix using the SVD of the estimated precoded channel
\begin{equation}
\hat{\mathbf{B}}_u[1,\nu]
=
\mathbf{U}_{\mathrm B,u}[1,\nu]
\mathbf{\Lambda}_{\mathrm B,u}[1,\nu]
\mathbf{V}_{\mathrm B,u}^{\mathrm H}[1,\nu].
\end{equation}
The first-stage combining matrix is then selected as
\begin{equation}
\mathbf{Q}_u[\nu]
= \mathbf{U}_{\mathrm{ B},u _{(:,1:N_{\mathrm c}})}[1,\nu],
\end{equation}
and the second-stage combining can be initialized as
\begin{equation}
\mathbf{W}_u[1,\nu]
=
\begin{bmatrix}
\mathbf{I}_{N_{\mathrm s}}\\
\mathbf{0}_{(N_{\mathrm c}-N_{\mathrm s})\times N_{\mathrm s}}
\end{bmatrix}.
\end{equation}

For $ 1<\tau \le \mathfrak{t}$, the first-stage combining matrix ${\mathbf{Q}}_u[\nu]$ remains fixed. Let $\hat{\mathbf{G}}_u[\tau,\nu]\in\mathbb{C}^{N_{\mathrm c}\times M}$ denote the estimated effective channel of UE $u$, defined as $\hat{\mathbf{G}}_u[\tau,\nu]$. Based on this effective channel, the BS computes the user-specific precoding matrix as
\begin{equation}
\mathbf{F}_u[\tau,\nu]
=
\eta_u[\tau,\nu]
\left(
\hat{\mathbf{G}}_u^{\mathrm H}[\tau,\nu]\hat{\mathbf{G}}_u[\tau,\nu]
+
\mu \mathbf{I}_M
\right)^{-1}
\hat{\mathbf{G}}_u^{\mathrm H}[\tau,\nu].
\label{eq:MU_peruser_MMSE_precoder2}
\end{equation} At UE $u$ side, the second-stage combining is updated from the estimated  precoded channel $\hat{\mathbf{D}}_u[\tau,\nu]$, where its SVD is written as
\begin{equation}
\hat{\mathbf{D}}_u[\tau,\nu]
=
\mathbf{U}_{\mathrm D,u}[\tau,\nu]
\mathbf{\Lambda}_{\mathrm D,u}[\tau,\nu]
\mathbf{V}_{\mathrm D,u}^{\mathrm H}[\tau,\nu],
\end{equation}
and the second-stage combining matrix is selected as
\begin{equation}
 {\mathbf{W}}[\tau,\nu] = \mathbf{U}_{\mathrm{D},u(:,1:N_\mathrm{s})}[\tau,\nu].
\end{equation}

{ 

\subsection{Achievable SE with Imperfect CSI for MU-MIMO}
\label{sec:uatf_mumimo}

We next provide an achievable downlink SE expression for the MU-MIMO system when the UEs rely on imperfect CSI obtained from pilot-based downlink estimation. We adopt the UatF bounding technique as described in Section~\ref{secCapacity}.

Consider subcarrier $\nu$ in coherence block $\tau$. UE~$k$ applies the two-stage combining matrix $\mathbf{U}_u[\tau,\nu]=\mathbf{Q}_u[\nu]\mathbf{W}_u[\tau,\nu]$ and forms the combined signal at UE $u$
\begin{equation}
\breve{\mathbf{y}}_u[\tau,\nu]
=
\mathbf{W}_u^{\mathrm H}[\tau,\nu]\mathbf{Q}_u^{\mathrm H}[\nu]\mathbf{y}_u[\tau,\nu].
\end{equation}
By inserting the MU-MIMO received signal model, we obtain
\begin{align}
\breve{\mathbf{y}}_u[\tau,\nu]
& =
\underbrace{\mathbf{G}_{u}[\tau,\nu]\mathbf{s}_u[\tau,\nu]}_{\text{desired signal}}
+
\underbrace{\sum_{i\neq u}\mathbf{G}_{i}[\tau,\nu]\mathbf{s}_i[\tau,\nu]}_{\text{multiuser interference}}\nonumber\\
& \quad
+
\underbrace{\mathbf{W}_u^{\mathrm H}[\tau,\nu]\mathbf{Q}_u^{\mathrm H}[\nu]\mathbf{n}_u[\tau,\nu]}_{\text{noise}},
\label{eq:uatf_received_mumimo}
\end{align}
where the effective channel is defined as
\begin{equation}
\mathbf{E}_{u}[\tau,\nu]
\triangleq
\mathbf{W}_u^{\mathrm H}[\tau,\nu]\mathbf{Q}_u^{\mathrm H}[\nu]\mathbf{H}_u[\tau,\nu]\mathbf{F}_u[\tau,\nu]
\in \mathbb{C}^{N_{\rm s}\times K}.
\end{equation}
 Define $\bar{\mathbf{E}}_{u}[\nu]
\triangleq
\mathbb{E}\!\left\{\mathbf{E}_{u}[\tau,\nu]\right\}$, where the expectation is taken with respect to the small-scale fading. The corresponding effective noise term is
\begin{align}
\tilde{\mathbf{n}}_u[\tau,\nu]
& \triangleq
\left(\mathbf{E}_{u}[\tau,\nu]-\bar{\mathbf{E}}_{u}[\nu]\right)\mathbf{s}_u[\tau,\nu] \nonumber\\
& \quad
+
\sum_{i\neq u}\mathbf{E}_{i}[\tau,\nu]\mathbf{s}_i[\tau,\nu]
+
\mathbf{W}_u^{\mathrm H}[\tau,\nu]\mathbf{Q}_u^{\mathrm H}[\nu]\mathbf{n}_u[\tau,\nu].
\end{align}
The covariance matrix of this effective noise is
\begin{align}
\mathbf{C}_{{\rm noise}_u}[\nu]
&\triangleq
\mathbb{E}\!\left\{\tilde{\mathbf{n}}_u[\tau,\nu]\tilde{\mathbf{n}}_u^{\mathrm H}[\tau,\nu]\right\}.
\label{eq:uatf_cov_mumimo}
\end{align}
Using \eqref{eq:uatf_received_mumimo}--\eqref{eq:uatf_cov_mumimo}, an achievable SE lower bound for UE~$u$ on subcarrier $\nu$ is given by
\begin{equation}
R_u^{\mathrm{UatF}}[\nu]
=
\log_2 \Big(\!\det\!\left(
\mathbf{I}_{N_{\rm s}}
+
\bar{\mathbf{E}}_{u}^{\mathrm H}[\nu]\,
\mathbf{C}_{{\rm noise}_u}^{-1}[\nu]\,
\bar{\mathbf{E}}_{u}[\nu] \Big)
\right).
\label{eq:uatf_rate_mumimo}
\end{equation}
Finally, the average downlink SE  is
\begin{equation}
\mathrm{SE}^{\mathrm{MU-MIMO}}
=
\frac{1}{S}\sum_{\nu=0}^{S-1}
\rho
\sum_{u=1}^{U} R_u^{\mathrm{MU-MIMO}}[\nu],
\label{eq:uatf_se_mumimo}
\end{equation}
where $\rho = 1- \frac{\mathsf{t}_\mathrm{p} + N_\mathrm{s}}{\mathsf{t}_c}$. The bound in \eqref{eq:uatf_rate_mumimo} is valid for arbitrary linear precoding and combining matrices. 
}

\section{Numerical Results}\label{sec:numerical}

{
In this section, we use Monte Carlo simulations to evaluate the SE of the proposed digital precoding and combining architecture. The aim is to demonstrate its performance advantages over conventional HBF schemes and alternative algorithms in both SU-MIMO and MU-MIMO scenarios, as well as to highlight the gains achieved by the proposed time-domain channel estimation compared to the frequency-domain approach.}

\subsection{Simulation Setup}

We simulate a mobile scenario where the UE travels along a linear trajectory. We consider that the BS is positioned at the 2D coordinates (2,5) meters and is equipped with a half-wavelength-spaced ULA comprising $M = 64$ antennas. The UE is equipped with a ULA containing $K = 16$ antennas. Both BS and UE's antennas are aligned parallel to the Y-axis. It begins its motion from coordinates (20,10) meters at a pedestrian speed of $v = 5$ m/s along a vertical line.\footnote{The results are also applicable to higher mobility scenarios. Increased UE velocities reduce both the beam and channel coherence time, which can be accommodated by correspondingly reducing the coherence block size.} The UE's location changes over time due to the movement, which creates realistic changes in the AoA and AoD, triggering the need for modifying the precoding and combining matrices. There are $N_{\mathrm{cl}} = 3$ clusters randomly located between the BS and UE, which are contributing to multipath propagation. The results are presented for $S = 512$ subcarriers. For path loss, we employ the 3GPP model outlined in \cite[Table 7.4.1-1]{3gpp2018study}, tailored for an urban microcell (UMi) environment and disregarding shadow fading. The carrier frequency is $f_\text{c} = 28$ GHz.   We assume $N_\mathrm{s} = 3$, $N_\mathrm{c} = 4$, $\mathsf{t}_{\mathrm{p}}= K$,  and $L=6$. In the next section, we present an average of achievable SE over various random cluster locations and fading realizations. As $\frac{512}{6}$ is not an integer, we pick 516 subcarriers for time-domain channel estimation first and then remove the last 4 subcarriers. 

\begin{figure}[!t]

        \centering \includegraphics[width=0.8\columnwidth]{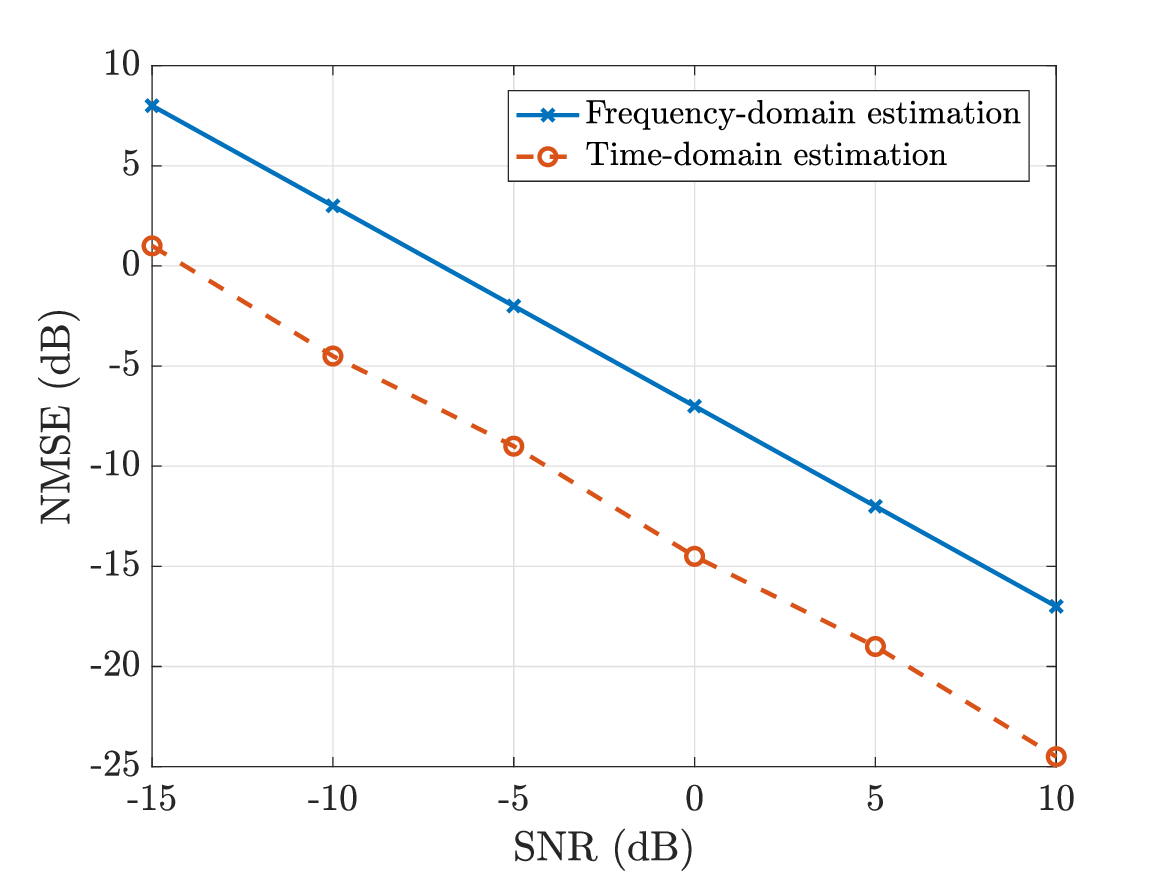}\caption{The NMSE vs SNR for frequency- and time-domain estimation of $\hat{\mathbf{H}}^{\mathrm T}[1,\nu]$.}
 \label{fig:tdfd}

\end{figure}

{

\subsection{Frequency-Domain and Time-Domain Estimation}

Fig.~\ref{fig:tdfd} shows the normalized mean squared error (NMSE) of the channel estimate $\hat{\mathbf{H}}^{\mathrm T}[1,\nu]$ as a function of the SNR for both frequency-domain and time-domain estimation. The SNR is defined based on the total transmit power normalized by the noise power. To ensure a fair comparison, the total pilot energy is kept constant across all methods. Hence, when fewer pilot subcarriers are used, the energy per pilot is increased proportionally. The NMSE is defined as $\mathrm{NMSE}
=
\frac{\mathbb{E}\!\left\{\|\hat{\mathbf{H}} - \mathbf{H}\|_{\mathrm F}^2\right\}}
{\mathbb{E}\!\left\{\|\mathbf{H}\|_{\mathrm F}^2\right\}}$
 and quantifies the relative estimation error.
Both methods exhibit an approximately linear decay in NMSE (in dB scale) with increasing SNR, which indicates that the estimation error is dominated by additive noise. However, the TD-based estimator consistently outperforms the FD approach over the entire SNR range, with a gain of roughly 6–8 dB. However, the TD-based estimator consistently outperforms the FD approach over the entire SNR range, with a gain of roughly 6--8\,dB. This performance advantage originates from the structural exploitation of the channel in the TD method. By leveraging the finite delay spread, the estimation problem is reduced from $S$ independent subcarrier coefficients to $L \ll S$ channel taps, which improves the conditioning of the estimation problem and enables implicit noise averaging across subcarriers. In contrast, FD estimation treats each subcarrier independently and does not exploit this structure, leading to higher estimation error. These results confirm that TD-based estimation provides more efficient reconstruction of $\mathbf{H}^{\mathrm T}[1,\nu]$.

\subsection{SU-MIMO System} 

\begin{figure}[!t]

        \centering \includegraphics[width=0.85\columnwidth]{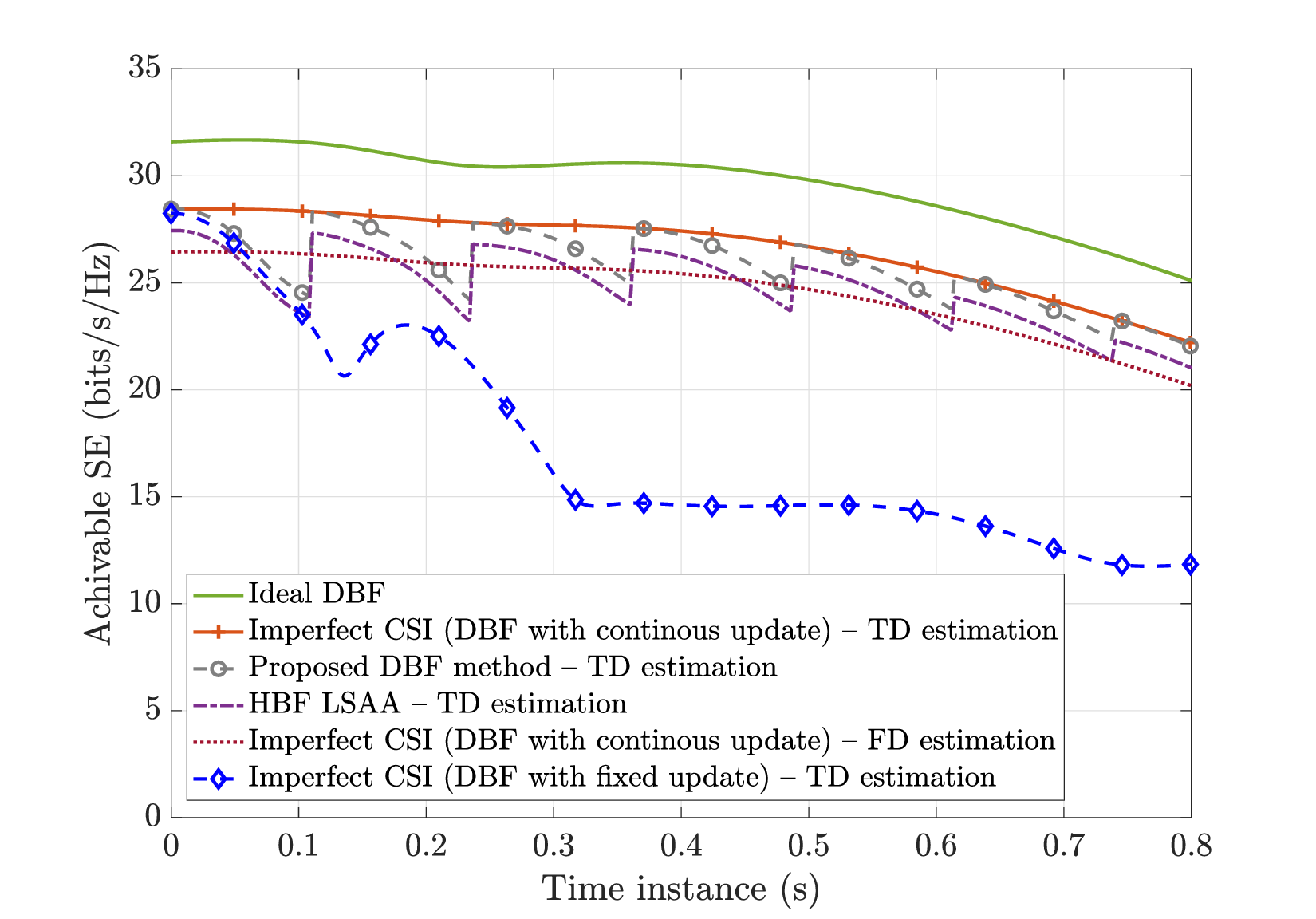}\caption{The average SE of SU-MIMO at different time instances when moving along a linear trajectory with speed $\upsilon=5$\,m/s.}
 \label{fig:singleuser}
\vspace{-3mm}
\end{figure}

Fig.~\ref{fig:singleuser} shows the achievable SE over time for SU-MIMO. The upper bound is given by “ideal DBF,” where perfect CSI is assumed and the first-stage combining matrix $\mathbf{Q}[\nu]$ is updated at every channel realization.
With imperfect CSI, the proposed DBF with continuous updates closely follows this bound, demonstrating that frequent updates of $\mathbf{Q}[\nu]$ enable effective tracking of channel variations. When $\mathbf{Q}[\nu]$ is kept fixed over the beam coherence time. In this simulation setup, the beam coherence time is $\mathsf{T}_\mathrm{B} = 102$,ms. A moderate performance loss is observed due to channel aging, but the degradation remains limited under the adopted beam coherence definition, noting that the exact impact depends on how this concept is defined (3 dB half-power beamwidth loss following \cite{khorsandmanesh2024beam}). We notice that the performance reduces when the channel changes, but the degradation from having a fixed first-stage combining is small, at most 14\%. This demonstrates that the hardware setup in Fig.~\ref{fig:hardware} can maintain high performance while reducing the computational complexity.
In contrast, fixing both $\mathbf{Q}[\nu]$ and $\mathbf{W}[\tau,\nu]$ leads to substantial performance loss, highlighting the importance of adaptive digital processing under mobility. The figure also compares time-domain (TD) and frequency-domain (FD) channel estimation. The TD-based estimation consistently provides higher SE, since it exploits the finite delay spread to save pilot resources and transmit power by a factor of $L/S$. This advantage is particularly visible over time, where TD estimation better preserves performance under channel variations, while FD estimation suffers from increased estimation errors and faster degradation. For comparison, we include SE results using the well-known HBF LSAA approach \cite{sohrabi2017hybrid}, where the first-stage combining is implemented using analog components and has a fixed configuration within the beam coherence time. Our proposed method demonstrates a clear performance advantage over LSAA. The bottommost curve depicts a scenario where both combining matrices, $\mathbf{Q}[\nu]$ and $\mathbf{W}[\tau,\nu]$, remain fixed throughout the simulation. In this case, the mobility significantly impacts the performance, showing substantial degradation over time.

\begin{figure}[!t]

        \centering \includegraphics[width=0.8\columnwidth]{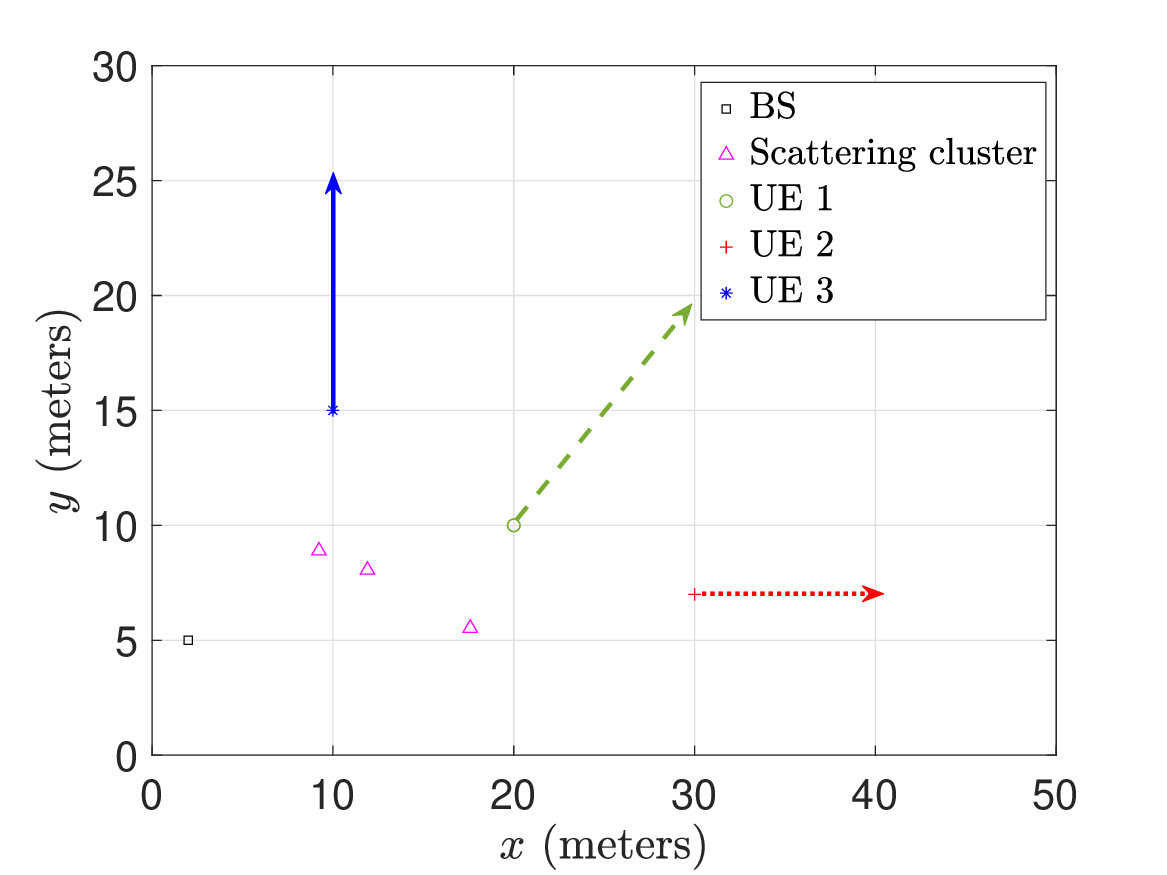}\caption{The 2D locations of the BS, UEs, their movement direction, and one example for scattering clusters.}
 \label{fig:location}
  \vspace{-3mm}
\end{figure}

\begin{figure}[!t]

        \centering \includegraphics[width=0.8\columnwidth]{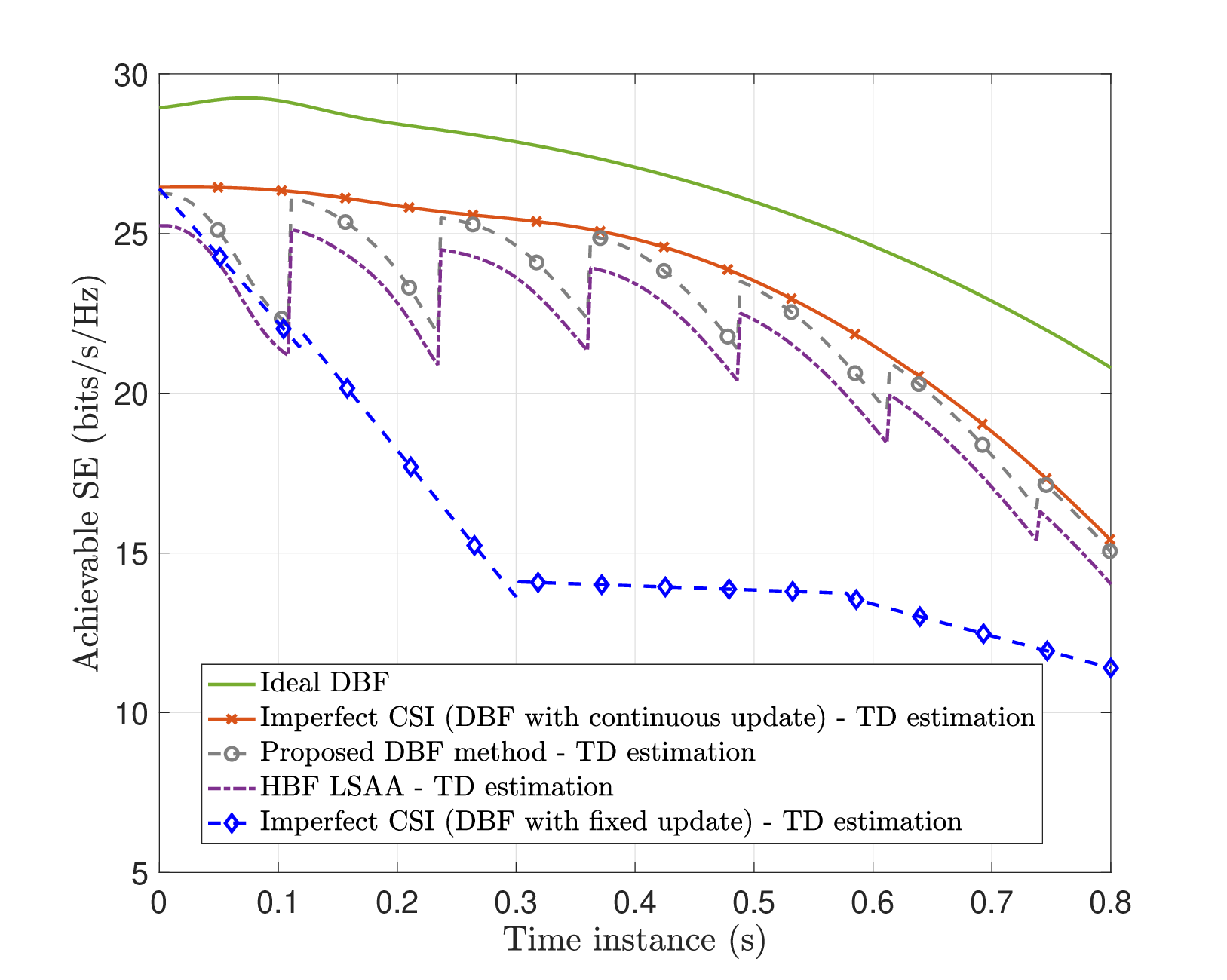}\caption{The average SE over time in a MU-MIMO system with three UEs, each moving at a speed of $\upsilon=5$,m/s. The results correspond to UE~3, as shown in Fig.~\ref{fig:location}}
 \label{fig:mumimo}
 \vspace{-4mm}
\end{figure}

\subsection{MU-MIMO System} 
Fig.~\ref{fig:location} illustrates one considered propagation scenario for the MU-MIMO setup, where multiple UEs are simultaneously served. This extends the previously considered single-user scenario. This includes the positions of the BS, one example of scattering clusters' locations, and three mobile UEs. The BS is located near the origin, while multiple scattering clusters are distributed in the environment to generate a spatially correlated mmWave channel. Each UE follows a predefined trajectory: UE~1 moves diagonally away from the BS, UE~2 moves horizontally, and UE~3 moves vertically starting from the position $(10,15)$ m. These mobility patterns induce time variations in both the large-scale geometry (angles and path gains) and the small-scale fading, which directly impact the channel estimation accuracy and beamforming performance. 

We focus on UE~3 and show the corresponding SE evolution in Fig.~\ref{fig:mumimo}. While the absolute SE is lower than in the single-user case due to inter-user interference and spatial multiplexing, the relative performance trends remain unchanged, confirming that the proposed framework extends effectively to the multi-user setting. The “Ideal DBF” provides an upper bound. With imperfect CSI, the proposed two-stage DBF with continuous updates closely approaches this bound, indicating efficient tracking of channel variations. The practical implementation incurs only a minor loss due to estimation errors but still clearly outperforms HBF schemes such as LSAA. The performance gap increases over time since hybrid architectures are more constrained in adapting to the instantaneous channel. Although the combining can be updated, the analog stage is frequency-flat and cannot capture the frequency selectivity of the channel, resulting in a suboptimal representation. As the channel evolves, this mismatch becomes more pronounced, leading to faster performance degradation than fully digital beamforming. The periodic drops in SE reflect channel aging between updates and are more evident for schemes with less accurate channel tracking. These results show the robust performance under mobility, also in multi-user scenarios.

\subsection{Digital Beamforming versus Hybrid Beamforming}
Fig.~\ref{fig:hybrid} shows the average SE versus the SNR for different combining schemes, evaluated at UE position (20,15) m (i.e., after 3 seconds). The proposed DBF with both TD and FD channel estimation is compared with HBF methods, including LSAA \cite{sohrabi2017hybrid}, PE-AltMin \cite{yu2016alternating}, and SS-SVD \cite{tsai2018sub}. The results show that the proposed DBF consistently outperforms all HBF schemes across the entire SNR range. This gain originates from the higher flexibility of fully digital combining, which enables more accurate adaptation to the instantaneous wideband channel conditions. Moreover, the TD-based channel estimation provides a systematic performance improvement over FD estimation, particularly at low and moderate SNRs, since it exploits the limited delay spread to reduce estimation noise and pilot overhead.
}

\section{Conclusions} \label{sec6}

HBF architectures suffer from practical limitations, including hardware inefficiencies and limited adaptability under mobility. To address these issues, we proposed a fully digital beamforming architecture with two-stage combining. The first stage performs dimension reduction prior to baseband processing, retaining the flexibility of digital processing while mitigating its implementation complexity. To enable efficient operation, we developed a pilot-based channel estimation framework and introduced a time-domain method that exploits the finite delay spread of OFDM channels. This approach improves estimation accuracy compared to conventional frequency-domain techniques. We further designed corresponding precoding and combining schemes and analyzed the achievable SE under imperfect CSI. The results show that updating the first-stage combining at the beam coherence timescale is sufficient to achieve near-optimal performance, even under mobility. Moreover, the proposed framework consistently outperforms HBF schemes and extends effectively to multi-user scenarios.

\begin{figure}[!t]

        \centering \includegraphics[width=0.8\columnwidth]{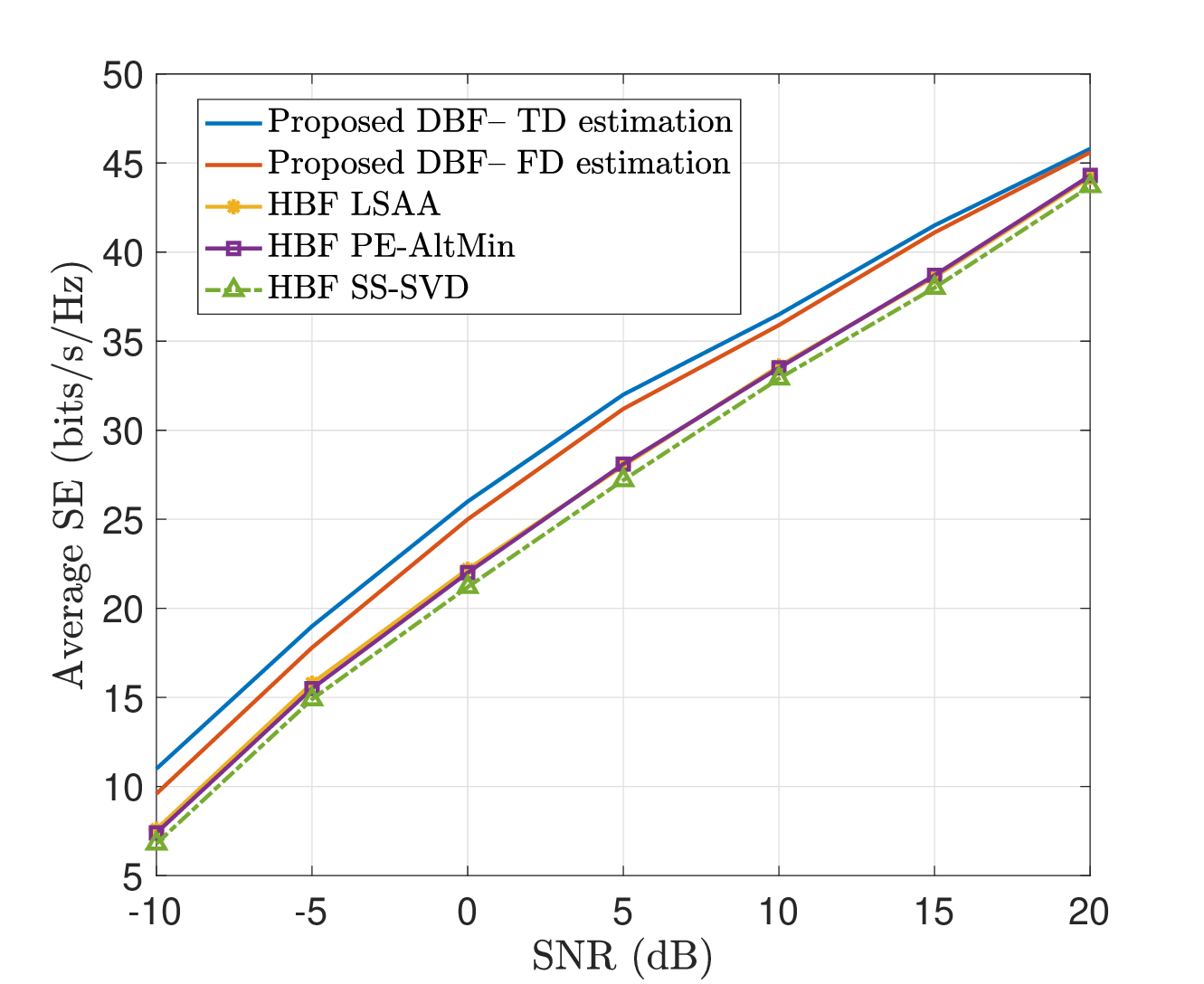}\caption{The average SE vs. the SNR of the proposed DBF method TD and FD estimation and different HBF schemes.}
 \label{fig:hybrid}
 \vspace{-3mm}
\end{figure}

\bibliographystyle{IEEEtran}
\bibliography{IEEEabrv,references}

\end{document}